\documentclass[pra,twocolumn,showpacs,showkeys,preprintnumbers,amsmath,amssymb,superscriptaddress]{revtex4}
\usepackage{graphicx}
\usepackage{dcolumn}
\usepackage{bm,amsmath,amsfonts,amssymb,mathrsfs}
\usepackage{epsfig}
\usepackage{color}

\newcommand{\mean}[1]{\langle #1 \rangle}

\newcommand{\ket}[1]{|#1\rangle}
\newcommand{\bra}[1]{\langle#1|}

\newcommand{\proj}[1]{\ket{#1}\bra{#1}}

\newcommand{\Def}[2]{\noindent{\bf Definition} {\bf #1.} {\it #2}}

\newcommand{\Le}[2]{\noindent{\bf Lemma} {\bf #1.} {\it #2}}
\newcommand{\Pro}[0]{\noindent{\bf Proof.}}

\newcommand{\be}[0]{\begin{equation}}
\newcommand{\ee}[0]{\end{equation}}
\newcommand{\bea}[0]{\begin{eqnarray}}
\newcommand{\eea}[0]{\end{eqnarray}}

\newcommand{\Tr}[1]{\text{Tr}\left [ #1 \right]}

	\newcommand{\tr}{\text{tr}}

\newcommand{\proofend}{$\Box$}

\usepackage[matrix,frame,arrow]{xy}
\usepackage{amsmath}
\newcommand{\idchannel}{\mathcal{I}}

\newcommand{\ud}{\mathrm{d}}

\begin{document}
\title{Schemes of transmission of classical information via quantum channels with many senders: discrete and continuous variables cases}

\author{L. Czekaj}
\email{lczekaj@mif.pg.gda.pl}
\affiliation{Faculty of Applied Physics and Mathematics,
Gda{\'n}sk University of Technology, 80-952 Gda{\'n}sk, Poland}
\affiliation{National Quantum Information Center of Gda\'nsk, 81-824 Sopot, Poland}

\author{J. K. Korbicz}
\affiliation{Institute of Theoretical Physics and Astrophysics,
  University of Gda\'nsk, 80-952 Gda\'nsk, Poland}

\affiliation{National Quantum Information Center of Gda\'nsk, 81-824
  Sopot, Poland}

\affiliation{Faculty of Applied Physics and Mathematics, Gda{\'n}sk
  University of Technology, 80-952 Gda{\'n}sk, Poland}

\affiliation{ICFO Institut de Cie`ncies Foto`niques, E-08860 Castelldefels, Barcelona, Spain}

\author{R. W. Chhajlany}

\affiliation{Faculty of Physics, Adam Mickiewicz University,
  Umultowska 85, 61-614 Pozna\'{n}, Poland}

\affiliation{Faculty of Applied Physics and Mathematics, Gda{\'n}sk
  University of Technology, 80-952 Gda{\'n}sk, Poland}

\affiliation{National Quantum Information Center of Gda\'nsk, 81-824
  Sopot, Poland}

\author{P. Horodecki}
\affiliation{Faculty of Applied Physics and Mathematics, Gda{\'n}sk
  University of Technology, 80-952 Gda{\'n}sk, Poland}

\affiliation{National Quantum Information Center of Gda\'nsk, 81-824
  Sopot, Poland}

\begin{abstract}
  Superadditivity effects in the classical capacity of discrete
  multi-access channels (MACs) and continuous variable (CV) Gaussian
  MACs are analysed. New examples of the manifestation of
  superadditivity in the discrete case are provided including, in
  particular, a channel which is fully symmetric with respect to all
  senders.  Furthermore, we consider a class of channels for which
  {\it input entanglement across more than two copies of the channels
    is necessary} to saturate the asymptotic rate of transmission from
  one of the senders to the receiver.  The 5-input entanglement of
  Shor error correction codewords surpass the capacity attainable by
  using arbitrary two-input entanglement for these channels.  In the CV
  case, we consider the properties of the two channels (a
  beam-splitter channel and a ``non-demolition'' XP gate channel)
  analyzed in [Czekaj {\it et al.}, Phys. Rev. A {\bf 82}, 020302 (R)
  (2010)] in greater detail and also consider the sensitivity of
  capacity superadditivity effects to thermal noise. We observe that
  the estimates of amount of two-mode squeezing required to achieve
  capacity superadditivity are more optimistic than previously
  reported.
\end{abstract}
\maketitle
\section{Introduction}

Quantum communication is a dynamically developing branch of quantum
information theory \cite{chuang_nielsen}. One of its central notions
is that of a quantum communication channel
\cite{chuang_nielsen,bennet_shor_QCC}, which models information
transfer from senders to receivers using quantum resources. The amount
of information which can be encoded in quantum states and reliably
sent through a quantum channel is measured, depending on the
communication scenario, by various channel capacities: (i) classical
capacity ($C$), defined as the maximal rate at which classical
information can be transmitted through the channel; (ii)
classical private capacity ($P$), which is the classical capacity
pertaining to the case when the transmitted bits are hidden from an
environment; (iii) quantum capacity ($Q$) characterizing the size of
the Hilbert space of states which can be  transmitted through
the channel. Quantum effects, associated with quantum channels, that
have recently attracted much attention are so-called ``activations''
and ``superadditivities''. For the quantum capacity $Q$, various
activations were based on bound entanglement, but the most spectacular
result was recently obtained in Ref.~\cite{smith_yard}, where an activation of
the type $0 \otimes 0 > 0$ was shown. In the case of private capacity
$P$, the corresponding superadditivity was found in
Ref.~\cite{winter_guo} (see also Ref.~\cite{smit_smolin_private}).
Quantum superadditivity of the classical capacity $C$ in the case of
Multiple Access Channels (MAC's) was shown in the
Ref.~\cite{discrete_mac} for discrete variables and in
Ref.~\cite{gaussian_mac} for continuous variables.  The question of
additivity of $C$ is still open for the one--sender one--receiver
scenario, although a substantial breakthrough on the superadditivity of
the Holevo function has recently been achieved in \cite{Hastings}.


In the present paper we study a variety of quantum multiple access
channels exhibiting superadditivity effects for classical capacity.
We do this for both discrete and continuous variable (CV) systems. In
particular, for the discrete variable case, we provide a new symmetric
scenario where both senders can benefit from capacity
superadditivity. This is in contrast to earlier examples studied in
Ref.~\cite{discrete_mac} where one of the senders only played a role
of an assistant with respect to the other fixed sender. We
  furthermore go beyond the standard dense coding protocol, which is
  based on two particle entanglement and present examples of channels
  where multipartite entanglement is required to achieve the optimum
  channel capacity. The use of multipartite entanglement can be seen
as the next step in the direction of optimization of the classical
capacity of quantum channels.  In particular it is shown that the 5-qubit
error correction codeword \cite{chuang_nielsen} entangled across 5 inputs beats any 2-input
based entanglement encodings for these channels.

In the CV context, we study the examples of Gaussian channels,
introduced in Ref.~\cite{gaussian_mac} in greater detail. We extend
the analysis of non-additive capacity regions and also study the
dependence of the classical capacity of the channels on the choice of
the set of input states. We show that for low energies, protocols
using two-mode entanglement surpass both coherent state and standard
single mode squeezed state encodings. Furthermore, we analyze the
sensitivity of the superadditivity effects to thermal noise and show
that the protocols are relatively sensitive to thermal noise or
losses in that 15 percent of power loss is sufficient to destroy the
effect.


The work is organized as follows. All necessary definitions are
introduced in Sec. \ref{background}.  Sections
\ref{discrete_additivity_theorem}-\ref{multiparticle_entanglement} are
devoted to the discrete variable case where we provide: a proof of
the classical additivity of capacity regions
(Sec. \ref{discrete_additivity_theorem}), an example of a symmetric
MAC, exhibiting superadditivity of the classical capacity
(Sec.~\ref{symmetrised_case}), an analysis of the influence of
multipartite entanglement on the capacity regularization
(Sec.~\ref{multiparticle_entanglement}) and an example of the
supperadditivity of regularized capacity regions
(Sec.~\ref{supperadditivity_of_reguralized_capaciy}). Continuous
variable MAC's are studied in Sections
\ref{locality_rule}-\ref{noise_and_realisation}, wherein: the locality
rule for continuous variable MAC's is presented
Sec.~\ref{locality_rule}, the dependence of the classical channel
capacities on the choice of input states is studied in
Sec.~\ref{strategies_comparison}), while the influence of thermal
noise is analyzed in Sec.~\ref{noise_and_realisation}).

\section{Basic definitions}\label{background}

The transmission of classical information through a quantum channel
corresponds to the following communication sequence
\cite{chuang_nielsen}:
\begin{equation}
x\mapsto\rho_x\mapsto\Phi(\rho_x)\mapsto\tr[\Phi(\rho_x) E_y]\mapsto y.
\end{equation}
The sender maps the message $x$ taken from some alphabet, into a state
$\rho_x$ of a quantum system, which in turn is sent through a quantum
channel $\Phi$ to the receiver.  The quantum channel models the
interaction of $\rho_x$ with the environment. It is assumed that none
of the users have access to the environment. The receiver obtains the
state $\Phi(\rho_x)$ and performs a measurement $\{E_y\}$ yielding
some output result $y$ from which he tries to infer the message sent
by the sender. The receiver knows the set of states $\{\rho_x\}$ as
well as the respective probabilities $p_x$ with which they are input
to the channel.  We distinguish two cases: (i) the states $\{\rho_x\}$
belong to a finite dimensional quantum space and $x$ is a discrete
variable (DV); (ii) the states $\{\rho_x\}$ are states of a bosonic
system and $x$ is a continuous variable (CV).  In the latter
situation, a restriction on the average energy sent through the
channel must be imposed to obtain a meaningful concept of channel
capacity, since cranking up the power of transmission indefinitely
allows perfect transfer of information.  The restriction usually takes
the form  of a constraint on the average photon number of the
input ensemble $\{p_x,\rho_x\}$: $\tr[\hat{N}\int p_x\rho_x \ud x]\leq
N$, where $\hat{N}$ is the photon number operator.

The sender may perform an encoding of his messages into code states to
reduce the probability that a message deciphered from the measurement
outcome disagrees with the one sent through the channel.  Code states
belong to the Hilbert space $\mathcal{H}^{\otimes n}$, describing the
input of $n$ copies of the channel $\Phi$, i.e. $\Phi^{\otimes n}$. As
$n\rightarrow\infty$ the probability of a decoding error can be made
arbitrary small.

The maximal rate at which information can be reliably transmitted
through a quantum channel is defined as its {\it classical capacity
  $C$}. By the well known result \cite{holevo}, the "single shot"
classical capacity $C^{(1)}(\Phi)$ is bounded by the {\it Holevo
  quantity}:
\begin{equation}
  C^{(1)}(\Phi)\leq\chi(\Phi)=\max_{\{p_x,\rho_x\}}\left( S(\Phi(\bar{\rho}))-\sum_x p_x S(\Phi(\rho_x))\right).
\label{holevo_capacity}
\end{equation}
where $\bar{\rho}=\sum_x p_x \rho_x$ is the mean input state and
$S(\rho)=-\tr[ \rho\log\rho]$ is the von Neuman entropy.
It can be shown that the above capacity can be achieved by product
code states over the copies of $\mathcal{H}$
(Holevo-Schumacher-Westmoreland coding theorem \cite{hsw}).

However, the input Hilbert space $H^{\otimes n}$ allows also for entangled states,
which may be useful for overcoming the above bound. This possibility
is quantitatively taken into account by considering the so-called {\it
  regularized classical capacity}:
\begin{equation}
  C^{(\infty)}(\Phi) = \lim_{n\rightarrow\infty}\frac{1}{n} \chi(\Phi^{\otimes n}).
\label{eq:regcapacity}
\end{equation}
The importance of considering entangled encodings is highlighted by
 Hastings' recent work ~\cite{Hastings}, who showed that there do
exist channels for which $C^{(\infty)}(\Phi)>\chi(\Phi)$.

In this paper we consider {\it multiple access channels} (MAC's),
where there are at least two senders (we will denote them as
$A,B,\ldots$, transmitting to one receiver $R$.  Each sender sends his
message independently of the other senders, i.e. their inputs are completely
uncorrelated.  They know only the input ensembles and agree upon a set
of rules governing the use of the channel: the first $n_1$ uses of the
channel consists of sending states from a fixed first ensemble, next
$n_2$ uses of the channel consist of states chosen from a second
ensemble and so on. This procedure is called time sharing
\cite{CoverThomas}.

For the case of two senders, a MAC acts as a mapping:
\begin{equation}
  \rho_{x_A}\otimes\rho_{x_B}\mapsto\Phi(\rho_{x_A}\otimes\rho_{x_B}).
\end{equation}
Here $x_A$ and $x_B$ are messages pertaining to senders $A$ and $B$
respectively.

The {\it capacity region} $\mathcal{R}(\Phi)$ of the classical MAC
$\Phi$ characterized by the conditional probability distribution $p(y_R|x_A,x_B)$
is defined as a set of vectors $R=\{R_A,R_B\}$ of rates, simultaneously
achievable by adequate coding and time sharing.  The capacity region
$\mathcal{R}(\Phi)$ of a classical two-sender MAC is given by the
convex hull of the rates $\{R_A,R_B\}$ for which there exist probability
distributions $p_{x_A},p_{x_B}$ of transmitted symbols and a joint
probability distribution $p_{x_Ax_By_R}=p(y_R|x_A,x_B)p_{x_A}p_{x_B}$
such that \cite{CoverThomas}:
\begin{eqnarray}
R_A &\leq& I(X_A:Y|X_B)\label{eq:cap_reg1}\\
R_B &\leq& I(X_B:Y|X_A)\\
R_A+R_B &\leq& I( X_A,X_B:Y)\label{eq:cap_reg3}.
\end{eqnarray}
where $I(X_A,X_B:Y)$ denotes the mutual information and $I(X_A:Y|X_B),
I(X_B:Y|X_A)$ are conditional mutual information quantities. These
quantities are related to the Shannon entropy $H(X)=-\sum_x p_x \log
p_x$ and conditional entropy $H(Y|X)=H(X,Y)-H(X)$ as follows:
$I(X:Y)=H(X,Y)-H(Y|X)$, $I(X:Y|Z)=H(X,Y|Z)-H(Y|X,Z)$. In the opposite way, for each vector of rates $R\in\mathcal{R}(\Phi)$ there exist input symbols probability distribution $p(x_A,x_B,Q)=p(x_A|Q)p(x_B|Q)p(Q)$ that following set of inequalities is fulfilled:
\begin{eqnarray}
R_A &\leq& I(X_A:Y|X_B,Q)\label{eq:cap_reg1q}\\
R_B &\leq& I(X_B:Y|X_A,Q)\\
R_A+R_B &\leq& I( X_A,X_B:Y|Q)\label{eq:cap_reg3q}.
\end{eqnarray}
Random variable $Q$ refers to time sharing procedure.


For the case of a quantum MAC $\Phi$ with two senders, a useful notion
is that of a ``classical-quantum'' state:
$\rho=\sum_{x_A,x_B}p_{x_A}p_{x_B}e_{x_A}\otimes
e_{x_B}\otimes\Phi(\rho_{x_A}\otimes\rho_{x_B})$ where
$\{e_{x_A}\}$,($\{e_{x_B}\}$) are projectors onto the standard basis  of
the Hilbert space controlled by sender $A$ ($B$) and
$\{p_{x_A},\rho_{x_A}\}$ ($\{p_{x_B},\rho_{x_B}\}$) is the ensemble of
code states of $A$ ($B$).

The single--shot capacity region $\mathcal{R}^{(1)}(\Phi)$ is
obtained as a convex closure of all rates $(R_A,R_B)$, for which there exist
 classical-quantum states $\rho$ fulfilling the following set of inequalities:

\begin{eqnarray}
  &R_A \leq I(X_A:Y|X_B)&\\
  &R_B \leq I(X_B:Y|X_A)&\\
  &R_T=R_A+R_B \leq I(X_A,X_B:Y).&
\end{eqnarray}
In distinction to the case of classical channels, the mutual
information is now given in terms of the von Neuman entropy
$I(X_A,X_B:Y) = S(\rho_{AB})+S(\rho_{R})-S(\rho_{ABR})$ and
$I(A:R|B)=\sum_{x_B} p_{x_B} I(A:R|B=\rho_{x_B})$. The Von Neuman
entropy is defined as $S(\rho)= -\tr [\rho \log\rho]$. $R_T$ denotes
the total capacity and is defined as $R_T=\sum_i R_i$.  In the
following, we will often refer to the notion of the regularized
capacity region $\mathcal{R}^{(\infty)}(\Phi)=\lim_{
  n\rightarrow\infty} \mathcal{R}(\Phi^{\otimes n})/n$.

Finally, we shall use the notion of {\it parallel composition} of
MAC's, which we illustrate here by an example of two classical
channels (denoted by $\Phi_I$ and $\Phi_{II}$) and two senders ($A$ and $B$). In
parallel composition sender $A$ has access to input ports $X_A^{I}$
$(X_A^{II})$ of the first (second) channel. $X_B^{I},X_B^{II}$ denote
input ports controlled by sender $B$.
For each input port $X_i^j$ there is a set of possible signals which
can be sent through the channel. The channels operate synchronously,
which means that communication process can be divided into steps. In
each step, user $A$ sends the vector of symbols
$x_A=\{x_A^{I},x_A^{II}\}$ while sender $B$ sends symbols
$x_B=\{x_B^{I},x_B^{II}\}$.  In each step a given channel is used by
every user exactly once. At the end of the communication step the
receiver obtains the output $y=\{y^{I},y^{II}\}$.

Let $p(y^{I}|x_A^{I}x_B^{I}), \big(p(y^{II}|x_A^{II}x_B^{II})\big)$ be the
transition probabilities for the MAC's $\Phi_I$ ($\Phi_{II}$), then the transition
probability for the parallel composition is given by:
\begin{eqnarray}
p(y|x_A,x_B)&=&p(\{y^{I},y^{II}\}|\{x_A^{I},x_A^{II}\},\{x_B^{I},x_B^{II}\})\\
&=&p(y^{I}|x_A^{I},x_B^{I})p(y^{II}|x_A^{II},x_B^{II}) \label{eq:prodchanprob}.
\end{eqnarray}
The parallel composition of quantum MAC's is defined as the straightforward generalization of the above concept.

\section{Quantum MACs in finite dimensional spaces}
\label{discrete_section}
\subsection{Additivity theorem for classical discrete multi access channels}
\label{discrete_additivity_theorem}

We shall state the additivity theorem for capacity regions
of classical discrete MACs in full generality.
First recall that the
capacity region $\mathcal{R}(\Phi)$ for a classical MAC with arbitrary
number of senders is given by the convex hull of the $\{R_i\}$ which
fulfill:
\begin{equation}
R_S\leq I(X_S:R|S^C)
\label{eq:gencapineq}
\end{equation}
where $S$ enumerates all subsets of senders and $R_S=\sum_{i\in S}
R_i$, while $S^C$ is the complement of the set $S$ \cite{CoverThomas}. For the 2-to-1 channels this reduce to the simple form of Eqs.~\ref{eq:cap_reg1}-\ref{eq:cap_reg3}.
The capacity region evaluated for fixed probability distribution of input symbols
$\tilde{p}=p(Q^I,Q^{II}) \prod_i p(X^I_i,X^{II}_i|Q^I,Q^{II})$ has the form (cf. Eq.~(\ref{eq:gencapineq})):
\begin{equation}
\tilde{\mathcal{R}}=\{ R\in\mathbb{R}^n: \forall_{S\subseteq E} R_S\leq  I(X_S:Y|X_{S^C},Q),\forall_{i\in E} R_i\geq 0\}.
\label{eq:fixprobdef}
\end{equation}

\begin{figure}
	\centering
		\includegraphics[scale=0.6]{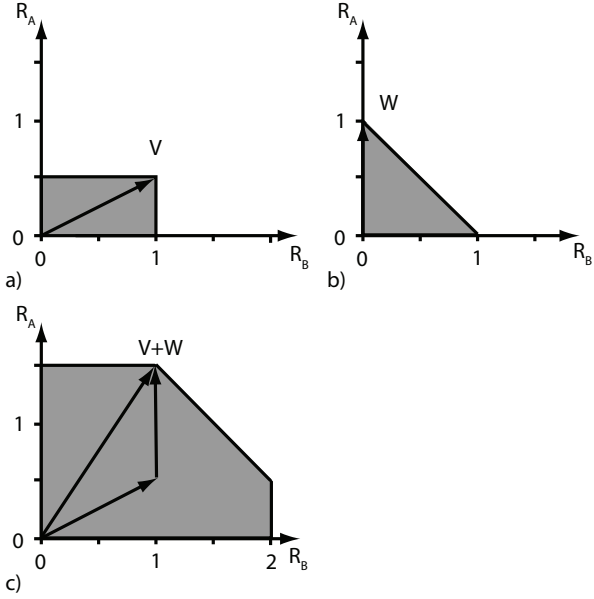}
		\caption{Additivity of the capacity regions for classical
          MACs. Capacity regions for channels $\Phi_{I}$ and
          $\Phi_{II}$ are presented respectively in Fig. a) and
          Fig. b). Capacity region of the parallel composition
          $\Phi_{I}\otimes\Phi_{II}$ of channels $\Phi_{I}$ and
          $\Phi_{II}$ is presented in Fig. c) and it is given by the
          geometrical sum of capacities regions from FIG. a) and
          FIG. b) (see Ref.~\cite{discrete_mac}).}
	\label{fig:geom_additivity}
\end{figure}

The \textbf{additivity theorem} states that the
achievable capacity region $\mathcal{R}$ of a channel being the
parallel composition of MACs is the geometrical sum of capacity
regions of the constituting channels. More formally, suppose $n$
MACs are used parallelly, with  each channel having $m$ senders. Let
$\tilde{R}=\{R_1,\ldots,R_m\}$ be the vector of achievable rates for
the composite channel, then the capacity additivity theorem
states that $\tilde{R}$ can be written as a sum of
vectors $\tilde{R}^{(j)}$ describing the capacity region of the $j$--th
MAC \cite{same_input_number_channel_extension}:
\begin{equation}
\mathcal{R}\left(\bigotimes_i\Phi_i\right)=\sum_i\mathcal{R}(\Phi_i)
\end{equation}
%
%
The additivity theorem for the case of channels with two senders is
graphically depicted  in FIG.~\ref{fig:geom_additivity}.

Here we prove only simple 2-to-1 scenario $\mathcal{R}(\Phi_{I}\otimes\Phi_{II})=\mathcal{R}(\Phi_I)+\mathcal{R}(\Phi_{II})$, complete prove will be postponed to appendix. We start with $(\subseteq)$. The outline is as follows: for arbitrary chosen vector of rates $\tilde{R}=(R_A,R_B)\in \mathcal{R}(\Phi_{I}\otimes\Phi_{II})$, by the capacity region definition we know that there exist probability distribution $\tilde{p}=p(X^I_A,X^{II}_A,X^I_B,X^{II}_B,Q^I,Q^{II})$ and corresponding fixed probability capacity region that $\tilde{R}\in\tilde{\mathcal{R}}(\Phi_{I}\otimes\Phi_{II})$.
We will use $\tilde{p}$ to construct probability distribution $\bar{p}=\tilde{p}_I\tilde{p}_{II}$ where $\tilde{p}_I=p(X^I_A,X^I_B,Q^I)$ ($\tilde{p}_{II}=p(X^{II}_A,X^{II}_B,Q^{II})$) is marginal probability distribution of input symbols of channel $\Phi_I$ ($\Phi_{II}$) obtained from $\tilde{p}$. Fixed probability capacity region corresponding to $\bar{p}$ will be denoted by $\bar{\mathcal{R}}(\Phi_I\otimes\Phi_{II})$. We will show that $\tilde{\mathcal{R}}(\Phi_I\otimes\Phi_{II})\subseteq\bar{\mathcal{R}}(\Phi_I\otimes\Phi_{II})\subseteq\mathcal{R}(\Phi_I\otimes\Phi_{II})$. Then we will show that
$\bar{\mathcal{R}}(\Phi_I\otimes\Phi_{II})=\tilde{\mathcal{R}}(\Phi_I)+\tilde{\mathcal{R}}(\Phi_{II})$ where $\tilde{\mathcal{R}}(\Phi_I)$ is fixed probability capacity region obtained for channel $\Phi_I$ for input symbols probability distribution $\tilde{p}_I$ and $\tilde{\mathcal{R}}(\Phi_I)$ has similar meaning for $\Phi_{II}$. From the relation $\bar{\mathcal{R}}(\Phi_I\otimes\Phi_{II})=\tilde{\mathcal{R}}(\Phi_I)+\tilde{\mathcal{R}}(\Phi_{II})$ we can draw conclusion that rates vector $\tilde{R}$ may be presented in the form $\tilde{R}=\tilde{R}_I+\tilde{R}_{II}$ where $\tilde{R}_{I(II)}\in\tilde{\mathcal{R}}(\Phi_{I(II)})\subseteq\mathcal{R}(\Phi_{I(II)})$, that finish the proof.

%

The following facts which will be usefull in futher considerations:
\begin{eqnarray}
H(Y|Q)&\leq&H(Y^I|Q^I)+H(Y^{II}|Q^I)\label{eq:2to1entropies1},\\
H(Y|X_B,Q)&\leq&H(Y^I|X^I_B,Q^I)\label{eq:2to1entropies2}+\\
&&H(Y^{II}|X^{II}_B,Q^{II}),\\
H(Y|X_A,X_B,Q)&=&H(Y^I|X_A^I,X^I_B,Q^I)\nonumber\\
&&+H(Y^{II}|X_A^{II},X^{II}_B,Q^{II})\label{eq:2to1entropies3}.
\end{eqnarray}
Eq.~\ref{eq:2to1entropies1} can be proved in following way:
\begin{eqnarray}
H(Y|Q)&=&\sum_{q}p(q)H(Y|Q=q)\\
&\leq&\sum_{\{q^I,q^{II}\}} p(\{q^I,q^{II}\})(H(Y^I|Q^I=q^I)\nonumber\\
&&+H(Y^{II}|Q^{II}=q^{II}))\label{eq:2to1proofCnd1}\\
&=&\sum_{q^I} p(q^I)H(Y^I|Q^I=q^I)\nonumber\\
&&+\sum_{q^{II}}p(q^{II}) H(Y^{II}|Q^{II}=q^{II}))\\
&=&H(Y^I|Q^I)+H(Y^{II}|Q^{II}),
\end{eqnarray}
where in Eq.~\ref{eq:2to1proofCnd1} we again make use of entropy subadditivity.
In similar way one can show Eq.~\ref{eq:2to1entropies2}.
To prove Eq.~\ref{eq:2to1entropies3} it is enought to observe, that conditional transition probability describing setup $\Phi_I\otimes\Phi_{II}$ factorizes (see Eq.~\ref{eq:prodchanprob}), hence we can write:
\begin{eqnarray}
H(Y|X_A,X_B,Q)\!\!&=&\!-\sum_{x_A,x_B,y,q}p\log p(y|x_A,x_B,q)\\
&=&\!-\!\!\!\!\sum_{x_A^I,x_B^I,y^I,q^I}\!\!\!\!p_I\log p(y^I|x_A^I,x_B^I)\\
&&\!-\!\!\!\!\!\sum_{x_A^{II},x_B^{II},y^{II},q^{II}}\!\!\!\!\!p_{II}\log p(y^{II}|x_A^{II},x_B^{II})\nonumber \\
&=&\!H(Y^I|X_A^I,X^I_B,Q^I)+\\
&&H(Y^{II}|X_A^{II},X^{II}_B,Q^{II})\nonumber.
\end{eqnarray}
where:
\begin{eqnarray} 
p&=&p(x_A,x_B,y,q)\\
&=&p(\{x_A^{I},x_A^{II}\},\{x_B^{I},x_B^{II}\},\{y^{I},y^{II}\},\{q^I,q^{II}\})\nonumber\\
p_I&=&p(x_A^I,x_B^I,y^I,q^I)\\
p_{II}&=&p(x_A^{II},x_B^{II},y^{II},q^{II}).
\end{eqnarray}

Now we are going to show that $\tilde{\mathcal{R}}(\Phi_I\otimes\Phi_{II})\subseteq\bar{\mathcal{R}}(\Phi_I\otimes\Phi_{II})$.
By the definition of capacity region, there exist input symbol probability distribution $\tilde{p}$ that the rates vector $\tilde{R}$ obeys Eqs.~\ref{eq:cap_reg1}-\ref{eq:cap_reg3}.
Using Eqs.~\ref{eq:2to1entropies1}-\ref{eq:2to1entropies3} we can bound RHS of Eqs.~\ref{eq:cap_reg1}-\ref{eq:cap_reg3} in following way:
\begin{eqnarray}
R_A&\leq&I(X_A:Y|X_B,Q)\\
&=&H(Y|X_B,Q)-H(Y|X_A,X_B,Q)\\
&\leq&H(Y^{I}|X^{I}_B,Q^I)+H(Y^{II}|X^{II}_B,Q^{II})\\
&&-H(Y|X_A,X_B,Q)\nonumber\\
&=&H(Y^{I}|X^{I}_B,Q^I)+H(Y^{II}|X^{II}_B,Q^{II})\\
&&-H(Y^{I}|X^{I}_A,X^{I}_B,Q^I)-H(Y^{II}|X^{II}_A,X^{II}_B,Q^{II})\nonumber\\
&=&I(X^{I}_A:Y^{I}|X^{I}_B,Q^I)+I(X^{II}_A:Y^{II}|X^{II}_B,Q^{II}),\label{eq:boundRA}
\end{eqnarray}
Analogical expression can be write for $R_B$.
\begin{eqnarray}
R_A+R_B&\leq&I(X_A,X_B:Y|Q)\\
&=&H(Y|Q)-H(Y|X_A,X_B,Q)\\
&\leq&H(Y^{I}|Q^I)+H(Y^{II}|Q^{II})\\
&&-H(Y|X_A,X_B,Q)\nonumber\\
&=&H(Y^{I}|Q^I)+H(Y^{II}|Q^{II})\\
&&-H(Y^{I}|X^{I}_A,X^{I}_B,Q^I)\nonumber\\
&&-H(Y^{II}|X^{II}_A,X^{II}_B,Q^{II})\nonumber\\
&=&I(X^{I}_A,X^{I}_B:Y^{I}|Q^I)\label{eq:boundRT}\\
&&+I(X^{II}_A,X^{II}_B:Y^{II}|Q^{II}).\nonumber
\end{eqnarray}
$I(X^{I}_A:Y^{I}|X^{I}_B,Q),I(X^{I}_B:Y^{I}|X^{I}_A,Q),I(X^{I}_A,X^{I}_B:Y^{I}|Q)$ are calculated for marginal distribution $\tilde{p}_I$ and $\tilde{p}_{II}$. Suming up, $\tilde{R}$ belongs to the region given by set of inequalities:
\begin{eqnarray}
R_A&\leq&I(X^{I}_A:Y^{I}|X^{I}_B,Q^I)\label{eq:sumcap1}\\
&&+I(X^{II}_A:Y^{II}|X^{II}_B,Q^{II})\nonumber\\
R_B&\leq&I(X^{I}_B:Y^{I}|X^{I}_A,Q^I)\\
&&+I(X^{II}_B:Y^{II}|X^{II}_A,Q^{II})\nonumber\\
R_A+R_B&\leq&I(X^{I}_A,X^{I}_B:Y^{I}|Q^I)\label{eq:sumcap3}\\
&&+I(X^{II}_A,X^{II}_B:Y^{II}|Q^{II})\nonumber.
\end{eqnarray}
These inequalities define region $\bar{R}(\Phi_I\otimes\Phi_{II})$ which is fixed probability capacity region for $\bar{p}$. We have shown capacity region inclusion.

We shall move to $\bar{\mathcal{R}}(\Phi_I\otimes\Phi_{II})=\tilde{\mathcal{R}}(\Phi_I)+\tilde{\mathcal{R}}(\Phi_{II})$.
Fixed probability capacity region $\tilde{\mathcal{R}}_I$ obtained  for input symbol probability distribution $\tilde{p}_I$ is given by:
\begin{eqnarray}
R_A&\leq&I(X^{I}_A:Y^{I}|X^{I}_B,Q^I)\\
R_B&\leq&I(X^{I}_B:Y^{I}|X^{I}_A,Q^I)\\
R_A+R_B&\leq&I(X^{I}_A,X^{I}_B:Y^{I}|Q^I).
\end{eqnarray}
Geometrical sum $\tilde{\mathcal{R}}_I+\tilde{\mathcal{R}}_{II}$ can be easy obtained as a convex hull of sums of vertices of the fixed probability capacity regions $\tilde{\mathcal{R}}_I,\tilde{\mathcal{R}}_{II}$ and is equal to the region $\bar{\mathcal{R}}$.
Because $\tilde{R}$ was chosen arbitrary, we have proven that $\mathcal{R}(\Phi_I\otimes\Phi_{II})\subseteq\mathcal{R}(\Phi_I)+\mathcal{R}(\Phi_{II})$.


$(\supseteq)$
Let $\tilde{R}_I\in\mathcal{R}(\Phi_I)$ belong to fixed probability capacity region with associated with input symbols probability $\tilde{p}_I$. Similar we have for $\tilde{R}_{II}$. It is easy to check by direct evaluation of Eq.~\ref{eq:cap_reg1q}-Eq.~\ref{eq:cap_reg3q}  that rates vector $\tilde{R}_I+\tilde{R}_{II}$ belongs to fixed probability capacity region of $\Phi_I\otimes\Phi_{II}$ obtained for input symbols probability distribution $\tilde{p}=\tilde{p}_I\tilde{p}_{II}$. That proofs $\tilde{R}_I+\tilde{R}_{II}\in\mathcal{R}(\Phi_I\otimes\Phi_{II})$.


\subsection{Supperadditivity}
Supperadditivity is defined as the situation when for a certain type
of capacity $\tilde{C}$ and two channels $\Phi_I,\Phi_{II}$, the
following holds:
\begin{equation}
\tilde{C}(\Phi_I\otimes\Phi_{II})>\tilde{C}(\Phi_I)+\tilde{C}(\Phi_{II}).
\label{eq:superadditivity}
\end{equation}
One may distinguish the following types of superadditivities: (a)
superaddivity of channel capacity, when $\tilde{C}=C^{(\infty)}$ (see
Eq. (\ref{eq:regcapacity})), (b) superadditivity of Holevo capacity,
when $\tilde{C}=\chi$, (c) self superadditivity, if $\tilde{C}=\chi$
and $\Phi_I=\Phi_{II}$.  For self superadditivity,
$C^{(\infty)}>C^{(1)}$.  Note that the  RHS of (\ref{eq:superadditivity}) expresses
the capacity achieved with product inputs
on $\Phi_I$ and $\Phi_{II}$. Superadditivity means that using encoded
states that are correlated (entangled) across uses of channels is
advantageous.

In the context of MACs superadditivity effects are identified in terms
of the capacity regions:
$\mathcal{R}(\Phi_I\otimes\Phi_{II})\varsupsetneq
\mathcal{R}(\Phi_I)+\mathcal{R}(\Phi_{II})$ where $+$ denotes the
geometrical sum of two regions. Superadditivity occurs if there exists
a vector in the region $\mathcal{R}(\Phi_I\otimes\Phi_{II})$ which
cannot be expressed as a sum of two vectors from $\mathcal{R}(\Phi_I),
\mathcal{R}(\Phi_{II})$ respectively.  To prove supperadditivty
effects in terms of the capacity regions it is enough to show that the
maximal rate achieved by one of the senders (say sender $A$)
exhibits superadditivity. This means that we may concentrate only on
the rate of a single sender or, in other words, show the effect
only by analysis of its ,,coordinate'' (or ,,dimension'') in the
multidimensional geometric regions
$\mathcal{C}(\Phi_I\otimes\Phi_{II})$, $\mathcal{C}(\Phi_I)$ and
$\mathcal{C}(\Phi_{II})$.

\subsection{Superadditivity effect in symmetric channels}
\label{symmetrised_case}
Examples of channels  presented
in \cite{discrete_mac,gaussian_mac}, which exhibit superadditivity effects, are highly unsymmetrical. One of the senders performs there a "remote" dense coding on the part of an entangled state transmitted by the other. In the described communication schemes one sender is a true
sender who transmits messages while the role of the others is only to help in the
communication process since their transmission rates is equal $0$.
It might suggest that in the channels based on the dense coding scheme
there is only a single super sender who takes advantage of the entangled state transmission. This is not the case as shown here. A channel can
be constructed that is symmetric with respect to the exchange of
senders facilitating a superadditivity effect for all of them.

\begin{figure}[h]
	\centering
	\includegraphics[scale = 0.3]{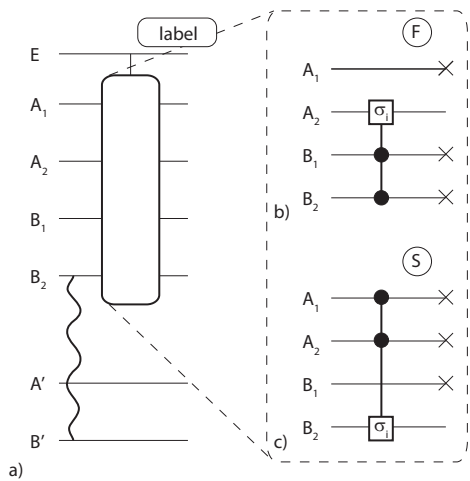}
\caption{Channel $\Phi$ from Sec.~\ref{symmetrised_case}. a) channel $\Phi$ working in parallel with identity channel $\idchannel$, waved line denotes entangled state, b) and c) two modes of work of channel $\Phi$.}
\label{fig:simetrised}
\end{figure}

Here we consider a channel $\Phi$ (see \ref{fig:simetrised}) with two senders: $A$ and $B$. Each of the senders controls two 1-qbit lines. The channel operates in two modes: $F$ and $S$. Each
occurs with probability $1/2$. In the first mode, the operation of the
channel is depicted in FIG. (\ref{fig:simetrised}.b). In the second
mode, $A$ and $B$ are swapped, {\it i.e.} lines $A_1$ and $A_2$
now belong to $B$ while $B_1$ and $B_2$ to $A$. The channel is
explicitly symmetric w.r.t. the senders.  Information that the first (second)
case occurred is sent to the receiver as a label $\ket{F} (\ket{S}$).
The cross at the end of lines denotes replacement of the transmitted
state by a completely mixed state. The action of the controlled
$\sigma_i$ gate is: $\proj{00}\otimes I + \proj{01}\otimes\sigma_x +
\proj{10}\otimes\sigma_z+\proj{11}\otimes\sigma_y$.
The capacity region $\mathcal{R}(\Phi)$ is upper bounded by the
following inequalities: $R_A\leq 1,R_B\leq 1,R_A+R_B\leq 1$, as a direct
consequence of the dimensionality of the output space (one-qubit
space).

Each user is supplied with an additional one qbit identity connection
with receiver.  These two channels will be jointly referred to as the
channel $\idchannel$. Note that its capacity region
$\mathcal{R}(\idchannel)$ is given by $R_A\leq 1, R_B\leq 1,
R_A+R_B\leq 2$.

\begin{figure}[h]
	\centering	
	  \includegraphics[scale=0.3]{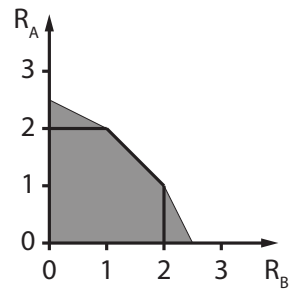}
      \caption{Lower bound for achieved capacity region for the
        channel $\Phi$ from Sec. \ref{symmetrised_case} working in
        parallel with identity channel. Thick lines refers to the
        upper bound for geometrical sum of capacity regions of
        component channels.}
\label{fig:SymCapReg}
\end{figure}

The upper bound for $\mathcal{R}(\Phi)+\mathcal{R}(\idchannel)$ thus
becomes $R_A\leq 2,R_B\leq 2, R_A+R_B\leq 3$.  On the other hand, the
lower bound for the achievable capacity region of the composite
channel $\mathcal{R}(\Phi\otimes\idchannel)$ can be seen in
FIG. \ref{fig:SymCapReg}.  To see this, we present a protocol which
achieves the capacity $(2.5,0)$. Due to symmetry of the channel, it
follows that the rates $(0,2.5)$ are also achievable. Notice
immediately that the rates $(1,2)$ and $(2,1)$ can be obtained by
product code states. All the other rates presented in
FIG. \ref{fig:SymCapReg} are obtained by time sharing.

Consider the following protocol: sender $A$ sends the
states $\ket{i}\ket{i'}$ with probability $1/8$ where
$\ket{i}\in\{\ket{00},\ldots,\ket{11}\}$ are all possible standard basis states of
two qubits, while $\ket{i'}\in\{\ket{0},\ket{1}\}$. The two-qubit states $\ket{i}$
are input to $\Phi$ while $\ket{i'}$ input to the supporting identity
channel $\idchannel$. $B$ sends the fixed state
$1/\sqrt{2}\ket{0}\left(\ket{00}+\ket{11}\right)$ with one qubit of
the Bell state sent through line $B_2$ and the other through the
supporting channel.

For given $\{i,i'\}$ the receiver gets
\begin{eqnarray}
  \rho_{i,i'} & = & \frac{1}{2}\proj{F}\otimes\frac{1}{2}\text{I}\otimes\proj{i_{A_2}}\otimes\frac{1}{8}\text{I}^{\otimes 3}\otimes\proj{i'}\nonumber\\
  & + & \frac{1}{2}\proj{S}\otimes\frac{1}{8}\text{I}^{\otimes 3}\otimes\proj{\psi_i}\otimes\proj{i'}, \nonumber
\end{eqnarray}
$\ket{F},\ket{S}$ denotes the mode of operation of channel $\Phi$. The
output state consists of the mode label and $6$ qbits.  The first $4$
qbits are output by the channel $\Phi$, while the $5$-th and
$6$-th qbits are outputs pertaining to $\idchannel$. If channel $\Phi$
works in mode 1, either the identity operation $I$ or $\sigma_x$ is
performed on the line $A_2$. However states sent by sender $A$
($\ket{0}$ and $\ket{1}$) are invariant under the mentioned operations
since the receiver obtains an unchanged state from  the line $A_2$.
If the channel $\Phi$ operates in mode 2, the controlled $\sigma_i$
gate fired by the state $|i \rangle $ from sender A, is performed on half of the Bell
state input by sender $B$. The result of this operation is denoted by
$\ket{\psi_i}$.  The entropy of the conditional output state $\rho_{i,i'}$ is
equal to $4.5$. Note that entropy has the same value for each input state $\ket{i}\ket{i'}$.

The mean output state is $\bar{\rho} = \frac{1}{8}\sum_{i,i'}\rho_{i,i'}$
and can be written as:
\begin{eqnarray}
  \rho & = & \frac{1}{2} \left( \proj{F}\frac{1}{64}\text{I}^{\otimes 6} + \proj{S}\frac{1}{64}\text{I}^{\otimes 6} \right) \\
  & = & \frac{1}{128}\text{I}^{\otimes 7}.
\end{eqnarray}
It has entropy  $S(\rho )=7$.
In presented scheme, sender $B$ transmits all the time the same state and attains a rate of $0$. Since the setup $\Phi\otimes\idchannel$ can be viewed as a channel with single sender $A$ while the state from the helper-sender $B$ is formally included to the environment.
By Holevo's theorem (see \ref{holevo_capacity}), we obtain that the
rate that sender $A$ can attain is thus $2.5$ bits.

Although rates $(2.5,0)$ and $(0,2.5)$ are achieved in the protocol where there is still one true sender while the other is helper-sender and there is no superadditivity of total rates $R_T=R_A+R_B$, potentially both of the senders can take advantage of entangled state transmission.

\subsection{Necessity of multi particle entanglement to approach
  regularized capacity region}

\label{multiparticle_entanglement}

In this section we give an example of a channel where senders must use
multiparticle entanglement states to achieve the regularized capacity
region.

\begin{figure}[h]
	\centering	
	  \includegraphics[scale=0.4]{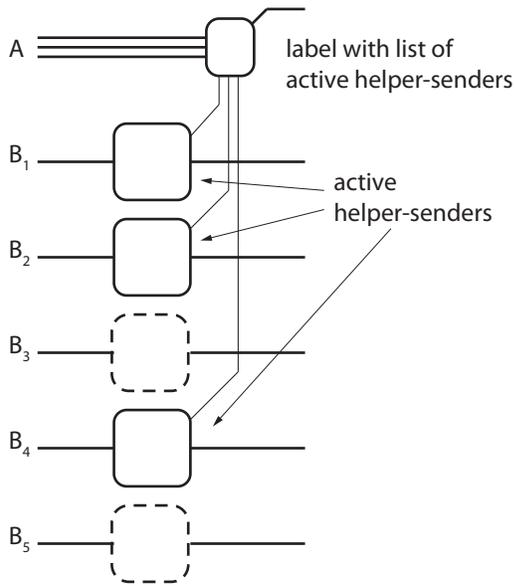}
      \caption{The channel described in
        Sec.~\ref{multiparticle_entanglement} with $n=5$
        helper-senders  and $n'=3$  lines belonging to sender
        $A$.  The message represented by the label
        $w={1,2,4}$ is additionally sent to the receiver which may be
        represented as a ,,flag'' state
        $|w\rangle=|001,010,100\rangle$.}
\label{fig:MultiUserSelect}
\end{figure}

We start by describing the class of channels $\Phi_{n,n'}$ that will be used in the
search for supperadditivity effects.  The channels have one
distinguished sender $A$ and $n$ helper-senders $B_i$. Sender $A$
controls $n'$ of 2-qbit lines which are measured in the standard basis by the channel (alternatively it can be seen as he controls $n'$ of 2-bit lines), senders $B_i$ control only 1-qbit lines. Each time the channel is used, one of the helper-senders is attributed to each 2-bit line of sender $A$. One helper-sender can be attributed only to one line of sender $A$.  Selected helper senders become active helper-senders.
It means that they participate in transmission of messages from sender $A$.
State from the active helper-sender is modified by the unitary operation from the set ${I,\sigma_x,\sigma_y,\sigma_z}$ which is triggered by the state of
the appropiate line of sender $A$.  States of the others helper-senders become unchanged.
Described selection of active helper-senders is performed in a random way.  Each selection can be chosen with equal probability.
States transmitted by $A$ are absorbed (i.e.  the
output degrees of freedom of $A$ are traced out).
The receiver obtains only the states coming from senders $B_i$ and a label $w$ with information about attribution of active helper-senders to lines of $A$.
For example if $n=3,n'=2$, the label $w=\{2,3\}$ tells receiver that  states
from senders $B_2$ and $B_3$ were chosen as the targets of the
unitaries controlled by first and second 2-qbit line of sender
$A$ respectively.    This channel is
schematically depicted in FIG. \ref{fig:MultiUserSelect}.  Note that
the message included in the label $w$ may be represented as a
$n'\left\lceil \log_2 n\right\rceil$-qubit state
$|w\rangle=|(i_1)_b,...,(i_{n'})_b\rangle$, where $i_k$ is the number
of the helper-sender chosen to be the target of the unitary operation
controlled by $k$-th line of sender $A$. $(.)_b$ denotes binary
representation of the value $i_k$. For example in the above mentioned
case of $n=3$ the label $w={2,3}$ corresponds to
$|w\rangle=|10,11\rangle$. We shall use this notation in the analysis
of a specific example.


\begin{figure}[h]
	\centering	
	  \includegraphics[scale=0.4]{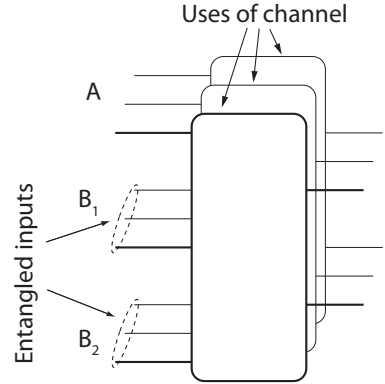}
      \caption{The parallel setup of channels described in
        Sec. \ref{multiparticle_entanglement}. The inputs of the
        channels used for transmission of entangled states are
        shown. The presented case consists of channels with two
        helper-senders $n=2$, each of which can send three particle
        entangled states $m=3$.}
\label{fig:ManyCopiesEnt}
\end{figure}

Here we study the parallel setup of $m$ copies of the channel $\Phi_{n,n'=1}$ from the
class described above. For simplicity we will denote the channel by $\Phi$. Please note that in following part of this section we choose $n'=1$. In the setup $\Phi^{\otimes m}$, senders $B_i$ can send
at most $m$-particle entangled states through their
inputs. Entanglement cannot be transmitted through inputs of two
different senders (see Fig.  \ref{fig:ManyCopiesEnt}).

We focus on the upper bound for the achievable rate for sender $A$.  We
restrict ourselves to the scheme where helper-senders $B_i$ send one state at all times. Vectors of rates for such
schemes take the form $(R_A,0)$. Formally we can consider the channel $\Phi^{\otimes
  m}$ in the setup as a $1$-to-$1$ channel and determine the  capacity
$C_A(\Phi^{\otimes m})$ of the sender $A$.  Now we prove that the
upper bound for the capacity $C_A(\Phi^{\otimes m})$ has the form:
\begin{equation}
C_A(\Phi^{\otimes m})\leq n\sum_{i=0}^m p^i(1-p)^{m-i}\binom{m}{i}\min(2i,m)
\label{ent_bound}
\end{equation}
where $n$ is the number of helper-senders, $m$ is the number of channels
used for transmission that is equivalent to the number of parties in the
entangled state pertaining to $B_i$, and $p=1/n$.

\textbf{Proof:} First we find an upper bound for the Holevo capacity of the
setup $\Phi^{\otimes m}$ in the case when the helper-sender $B_i$ was active $l_i$ times. Then we use these
results to calculate the upper bound for the capacity of $\Phi^{\otimes m}$.

The orthogonal label $\ket{w_j}$ describes which sender $B_i$ was active in the $j$
copy of $\Phi$. Label
$\ket{w}=\ket{w_1,\ldots,w_m}=\ket{w_1}\otimes\ldots\otimes\ket{w_m}$
is the complete list of the active helper-senders in the setup. Given the label we know
that sender $B_i$ was active $l_i$ times.
The probability of occurrence of the situation described in $\ket{w}$
is given by $p_w=p^m$.

\begin{figure}
	\centering
		\includegraphics[scale=0.4]{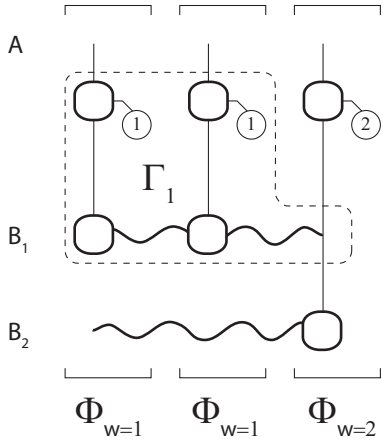}
	\caption{\label{fig:phi_gamma_equivalence}
	Equivalence $\Phi^{\otimes m}_w= \otimes_i \Gamma_i$ on example of channel $\Phi^{\otimes 3}_{w=\{1,1,2\}}$. Dashed line delimits channel $\Gamma_1$, waved lines denotes entangled states from by senders $B_1$ and $B_2$, label with active helper-sender is in the circle.}
\end{figure}

Suppose that $\ket{w}$ is obtained as the result of $\Phi^{\otimes
  m}$. This fixes the attribution of senders $B_i$ to the lines of
$A$. We denote this case as $\Phi^{\otimes m}_w$. Now, the $m$ uses of
the channel $\Phi$ can be thought as $n$ separate channels $\Gamma_i$
$m$.  The input of each channel $\Gamma_i$ consists
of the subset of lines from $A$ and all lines from $B_i$. None of the
$\Gamma_i$ share input lines with any other $\Gamma_j$.
Each channel $\Gamma_i$ has $2 l_i$ qbits input
from sender $A$, $m$ qbits input from sender $B_i$ and a $m$ qbit
output. The equivalence $\Phi^{\otimes m}_w= \otimes_i \Gamma_i$ is depicted in FIG.~\ref{fig:phi_gamma_equivalence}.

Taking into account dimensionality, one can infer that $A$ can
transmit  at most $\min (2 l_i,m)$ classical bits of
information through $\Gamma_i$. Given $\ket{w}$, the channels $\Gamma_i$ work
independently. There cannot be entanglement shared between $\Gamma_i$
and $\Gamma_j$ because sender $A$ transmits only classical states and
users $B_i$ and $B_j$ cannot share entanglement due to definition of
MAC. This leads to the total conditional capacity:
\begin{equation}
C_A (\Phi^{\otimes m}_w) = \sum_{i=1}^n C(\Gamma_i) = \sum_i \min (2 l_i,m)
\label{chiaw}
\end{equation}

The following observation is helpful for the  calculation of
$C_A(\Phi^{\otimes m})$. Consider channel $\Delta(\rho)=\sum_w
p_w\Delta_w(\rho)\proj{w}$ which acts with probability $p_w$ as
channel $\Delta_w$. Assume again that the  label $\ket{w}$ is sent to
the receiver which identifies the case that occurs. For this channel we have:
\begin{eqnarray}
  C(\Delta)&=&
  \max_{\{p_x,\rho_x\}} S\left(\Delta\left(\sum p_x \rho_x \right)\right)
  -\sum_x p_x S\left(\Delta\left( \rho_x \right)\right) \label{eq:thsumcaps} \\
  &=&\max_{\{p_x,\rho_x\}} S\left(\sum_w p_w\Delta_w\left(\sum p_x \rho_x \right)\proj{w}\right)\nonumber\\
  &&-\sum_x p_x S\left(\sum_w p_w\Delta_w\left( \rho_x \right)\proj{w}\right)\\
  &=&
  \max_{\{p_x,\rho_x\}}\sum_w  p_w \left\{ S\left(\Delta_w\left(\sum p_x \rho_x \right)\right)+H(\{p_w\})\right.\nonumber\\
  &&-\left.\sum_x p_x S\left(\Delta_w\left( \rho_x \right)\right)-H(\{p_w\})\right\}\\
  &\leq&
  \sum_w  p_w \max_{\{p_x^w,\rho_x^w\}} \left\{ S\left(\Delta_w\left(\sum p^w_x\rho^w_x \right)\right)\right.\\
  &&\left.-\sum_x p^w_x S\left(\Delta_w\left( \rho^w_x \right)\right)\right\}\nonumber\\
  &=&\sum_w p_w C(\Delta_w) \label{eq:thsumcape},
\label{chi:linearcomb}
\end{eqnarray}
where equality occurs if the same ensemble achieves the capacity of each channel $\Delta_w$. Similar argumentation can be use to show that rates achieved for channel $\Delta$ in certain protocol obey:
\begin{equation}
R(\Delta)=\sum_w p_w R(\Delta_w)\label{eq:ratesSum},
\end{equation}
where $R(\Delta_w)$ are the rates achieved by this protocol in case of $\Delta_w$.


Using the above observation, and substituting $\Delta = \Phi^{\otimes
  n}_w$, $p_w = p^m$ and bound $C(\Delta_w)$ by $C_A(\Phi^{\otimes
  m}_w)$ (\ref{chiaw}), we obtain:
\begin{eqnarray}
C_A(\Phi^{\otimes m})&\leq& p^{m} \sum_w \left( \min(2 l_1(w),m)+\right.\\
&&\left.\ldots+\min(2 l_n(w),m) \right),\nonumber\\
\end{eqnarray}
where $l_i(w)$ denotes value of $l_i$ encoded in label $w$.
After some rearrangement:
\begin{eqnarray}
C_A(\Phi^{\otimes m})&\leq& p^{m} \sum_w \left( \min(2 l_1(w),m)+\ldots\right)\\
&=& p^{m} \sum_{l_1+\ldots+l_n=m} \frac{m!}{l_1!\cdot\ldots\cdot l_n!} \label{formulal2}\\
&&\left( \min(2 l_1,m)+\ldots+\min(2 l_n,m)\right)\nonumber\\
&=&n p^{m}\! \sum_{l=0}^m \binom{m}{l} \min(2 l,m) \alpha_l \label{formulal2a}\\
&=&n\sum_{l=0}^m\!\binom{m}{l}\!\min(2 l,m)p^{m}(n\!-\!1)^{m-l} \label{formulal2b}\\
&=&n\!\sum_{l=0}^m p^l(1-p)^{m-l}\binom{m}{l}\!\min(2l,m)\label{formulal2c}.
\end{eqnarray}
In Eq.~(\ref{formulal2}) we collected in the common factor all
$w$ with the same $\{l_1,\ldots,l_n\}$. Because formulas with
$\min(l_2,m),\ldots,\min(l_n,m)$ in Eq.~(\ref{formulal2}) have the
same form as the one with $\min(l_1,m)$, we omitted them and introduced
in Eq.~(\ref{formulal2a}) factor $n$. Moreover we introduced
$\alpha_{l} = \sum_{l_2+\ldots+l_n=m-l}\binom{m-l}{l_2,\ldots,l_n}$.
In Eq.~(\ref{formulal2b}) we used the relation
\begin{equation}
\sum_{k_1+\ldots+k_n= m}\binom{m}{k_1,\ldots,k_n}= n^m.
\end{equation}
Recalling that $p=1/n$ leads the relation $(n-1)p=(n-1)/n=1-p$
which was used in Eq.~(\ref{formulal2c}).

\begin{figure}[h]
	\centering	
    \includegraphics[scale=0.7]{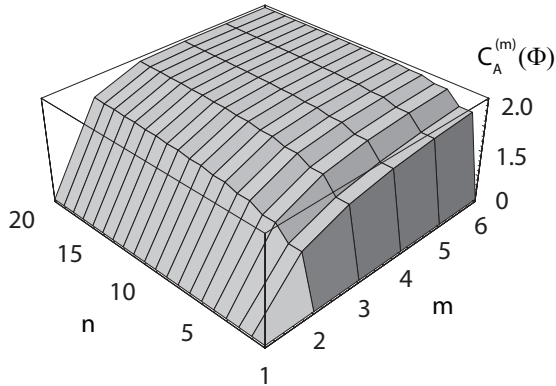}
	\caption{Upper bound for reguralized capacity $C^{(m)}_A=\frac{1}{m}C_A(\Phi^{\otimes m})$ as a function of $m$ - number of channel copies the capacity is evaluated on - and number of helper sender $m$ in the channel $\Phi$.}
\label{fig:asymp_cap}
\end{figure}

The upper bound given by (\ref{ent_bound}) is achieved in the case of
$m\in\{1,2,5\}$ by the protocols which runs as follows: $A$ transmits with equal probability all
states from the standard basis of his $2 m$-qbit input space of
$\Phi^{\otimes m}$ while all $B_i$'s transmit  either state $\ket{0}$ from the standard base, one of the
Bell states $\ket{\Phi^+}$ or $\ket{0_L}$ -- the $5$ qbit correction code word
(see \cite{chuang_nielsen}) for $m=1$, $m=2$ or $m=5$ case respectively.

\begin{eqnarray} \ket{0_L}&=&\frac{1}{4}\big[\ket{00000}+\ket{10010}+\ket{01001}+\ket{10100}\\
  &&+\ket{01010}-\ket{11011}-\ket{00110}-\ket{11000}\nonumber\\
  &&-\ket{11101}-\ket{00011}-\ket{11110}-\ket{01111}\nonumber\\
  &&-\ket{10001}-\ket{01100}-\ket{10111}+\ket{00101}\nonumber\big]
\end{eqnarray}


To prove this, we will show that for each $w$, an ensemble used in the protocol gives  (\ref{chiaw}).
Equality in Eq.~\ref{ent_bound} then becomes a simple consequence of relation Eq.~\ref{eq:ratesSum}.

Assuming knowledge of $w$, the output entropy of the channel is equal
to $0$
for each state transmitted in described protocol. Hence we have to
check if, under condition of $w$, the mean output state entropy
$S(\rho_w)$ reaches $\sum_i\min(2 l_i(w),m)$. Since senders $B_i$ are
uncorrelated, we can focus only on sender $B_i$ and consider only
value $l_i$. Let the set $e$ contain all positions where the state
coming from sender $B_i$ was affected by the channel $\Phi$
($|e|=l_i$). We denote by $\mathcal{E}^{k}(\rho)$ the completely
depolarizing channel acting on $k$-th qbit of the state $\rho$. For
given $e$, part of the mean output state coming from $B_i$ has the form
$\rho_{e}=\left(\bigotimes^{l_i}_{j=1}\mathcal{E}^{(e_j)}\right)[\ket{\phi}\,\bra{\phi}]$
where $e_j$ denotes the $j$-th element of $e$ and $\ket{\phi}$ is $\ket{0}$, $\ket{\Phi^+}$ or $\ket{0_L}$. The condition of whether  $S(\rho_{e})=\min(2 l_i ,
m )$ occurs for all $e$ was checked numerically. The program enumerated all
$e$, then for each $e$ it computed state $\rho_e$ and its entropy
$S(\rho_e)$. Obtained results confirmed that $S(\rho_{e})=\min(2 l_i ,
m )$ for $\ket{0}$, $\ket{\Phi^+}$ and $\ket{0_L}$.



In the presented protocol, entanglement increases diversity of the
mean output state. An important feature of the code state from the $5$
qbit correction code is that the increase of entropy of the output state
depends only on the number of qbits affected by the unitary. It does
not depend on localization of affected qbits. We cannot exceed $m$
bits of entropy per state hence the closer $l_i$ is to $m$ the smaller
is the entropy increase. Due to the asymptotical equipartition
property, for $n>1$ the larger the entanglement in the state, the
smaller is the chance that $l_i$ will be close to the $m$.

The above analysis opens the possibility of further analysis
concerning type of entanglement is the best in case of various
channels. The possible classification of noise with respect to classes
of entanglement seems especially interesting, for instance one can ask
whether there are any channels for which cluster type entanglement is
the best in saturating the asymptotic rates of the channel.  We leave
these type of questions for further research.


\subsection{Supperadditivity of regularized capacity.}
\label{supperadditivity_of_reguralized_capaciy}

We now turn to the study of the supperadditivity effect for
regularized capacity.  We will investigate a setup which consists of
 two channels of the type already described in
Sc.~\ref{multiparticle_entanglement}. For the channel $\Phi_I$ we choose
$n=10,n'=9$ and for the channel $\Phi_{II}$ we choose $n=10,n'=1$.  We
are interested in maximal transmission rate from sender $A$, that is
the case when all senders $B_i$ help sender $A$ by transmitting the
same states all the time. Their rates are equal $0$. Formally we can include senders $B$ to the environment and view channels $\Phi_I$ and $\Phi_{II}$ as $1$-to-$1$ channels.

First we show that for channel $\Phi_I$ and $\Phi_{II}$, upper bounds
$C_A$ for rates achievable by sender $A$ fulfill $C_A^{(\infty)}>
C_A^{(1)}$. For this, we consider a protocol where senders $B_i$
transmits one of the Bell states and show that this protocol achieves
a regularized rate strictly greater than $C_A^{(1)}$.  Calculation
will be performed for general sizes of the set of selected
helper-senders equal $n'$.  The single shot capacity is given by the
joint dimensionality of states of the selected helper-senders and it
reads: $C_A^{(1)}=n'$. In case of two uses of the channel and Bell states transmission, the
probability that the same set of selected helper-senders was chosen
twice is $p=1/\binom{n}{n'}$. With probability $1-p$ sets of the
selected helper-senders in the first and second uses of the channel
differ in at least one sender $B_i$. It means that input Bell state of two sender, lets say $B_1$ and $B_2$, was affected by the channel only once and, due to dense coding, the states carry full information from appropriate 2-qbit input lines of $A$ (line $1$ and $2$). In this case sender $A$ take advantage of transmission of additional $2$ bits of information.
Under condition of output label $w$, output entropy of the
channel is $0$ therefore the rate achievable by the protocol for given $w$ is equal to
entropy of the mean output state (strictly speaking entropy of the part coming from senders $B$). As usual, sender $A$ transmits with
equal probability all states from the standard basis.
Recalling to Eq.~\ref{eq:ratesSum}, the rate achievable by sender $A$ is at least
$R_A= p 2n' + (1-p)(2n'+2) = 2(n'+(1-p))$. This leads to $C_A^{(\infty)}\geq (1/2)
R_A=n'+(1-p)>n'=n'\geq C^{(1)}$.

Now we pass to the supperadditivity of the regularized capacities. We
again refer to (\ref{eq:superadditivity}).  We will show that
$C_A^{(1)}(\Phi_{I}\otimes\Phi_{II})>C_A^{(\infty)}(\Phi_{I})+C_A^{(\infty)}(\Phi_{II})$.
We first provide upper bounds for $C_A^{(\infty)}(\Phi_{I})$ and
$C_A^{(\infty)}(\Phi_{II})$.

Recall that in this situation, entangled states can be transmitted only through the inputs controlled by the same users. Channel capacity is upper bounded by the minimum value of logarithm of its input and output spaces. Therefore for the channel $\Phi_{n,n'}$, we have $C_A^{(m)}\leq 1/m \min(2n' m,n m) = \min(2n',n)$ and it leads to $C_A^{(\infty)}(\Phi_I)\leq \min( 2\times 9 , 10 ) = 10$ and $C_A^{(\infty)}(\Phi_{II})\leq \min( 2\times 1 , 10 ) = 2$.


Now we move to the case $\Phi_I\otimes\Phi_{II}$, i.e. the case where entanglement between inputs of channels $\Phi_I$ and $\Phi_{II}$ controlled by the same user
is allowed. One can use the following protocol to provide a lower bound for
$C^{(1)}_A(\Phi_{I}\otimes\Phi_{II})$: sender $A$  only uses inputs of
$\Phi_I$, and sends each state from the standard basis of the input
space of $\Phi_I$ with the same probability; through channel
$\Phi_{II}$, he sends only one chosen state $\ket{00}$ all the time.
It is easy to see that channel $\Phi_{II}$ does not change the states coming from senders $B_i$ and in fact it can be seen as an identity channel.
Senders $B_i$ send one
chosen Bell state $\ket{\Phi^+}$. The first qbit of the Bell state goes through the
channel $\Phi_I$ while the second through the channel $\Phi_{II}$.  If
the qbit is affected by the channel $\Phi_I$, the dense
coding scheme is reproduced. Each time the setup $\Phi_I\otimes\Phi_{II}$ is used, all the lines controlled by $A$ find as a target different Bell states. Therefore rate achieved by protocol is given by the dimensionality of input space of channel $\Phi_{I}$ controlled by sender $A$ and reads $18$ bits. It is lower bound for $C_A^{(1)}(\Phi_{I}\otimes\Phi_{II})$ and shows that
$C_A^{(\infty)}(\Phi_{I}\otimes\Phi_{II})\geq C_A^{(1)}(\Phi_{I}\otimes\Phi_{II})\geq 18>12\geq C_A^{(\infty)}(\Phi_{I})+C_A^{(\infty)}(\Phi_{II})$ and proofs that
supperadditivity effect indeed occurs.


\section{Quantum Gaussian MACs}
\label{cv_section}

We shall now consider the capacity properties of Gaussian Multi-Access
channels. Before going further, we first collect certain basic notions
and definitions that will be subsequently useful.

Recall first the concept of classical Gaussian multiple access
channels \cite{CoverThomas}.  Inputs and outputs of classical CV
gaussian MACs are real numbers. The gaussian MAC models the influence
of additive gaussian noise $Z$ (with variance $S$) on the total input
signal, {\it i.e.} the output is
\begin{equation}
Y = \sum_i X_i + Z
\end{equation}
To prevent unphysical infinite capacities, the power constraints are
imposed on the input signals $\mean{X^2_i}\leq P_i$. Under these
constraints, the capacity
region for the classical gaussian MAC channel is given by \cite{CoverThomas}:
\begin{equation}
\sum_i R_i \leq C(\sum_i P_i /S)
\label{classcapreg}
\end{equation}
where $C(x) = 1/2 \log (1+x)$.

For a {\it quantum} Gasussian MAC, the input and the output spaces are
described by infinite dimensional Hilbert spaces, isomorphic to those
describing a finite number of bosonic modes \cite{Hayashi07PR}. The latter
are equipped with the ``position'' and ``momentum'' canonical
observables
$\{\hat{x}_1,\ldots,\hat{x}_n,\hat{p}_1,\ldots,\hat{p}_n\}$ fulfilling
the commutation rules $[\hat{x}_i,\hat{p}_j] = \mathrm{i}
\delta_{i,j}$, where $i,j$ enumerate modes of the system. States of a
bosonic system can be expressed in terms of characteristic functions
$\chi_\rho(\xi)=\Tr{\rho W_\xi}$ where $W_\xi=\exp(-\mathrm{i} \xi^T
R)$ is the so-called Weyl operator and
$\hat{R}=(\hat{x}_1,\hat{p}_1,\ldots,\hat{x}_n,\hat{p}_n)^T$ is the
vector of canonical observables \cite{Hayashi07PR,introquantopt}.
Gaussian states are the states whose characteristic functions are
gaussian:
\begin{equation}
\chi(\xi)=\exp\left[-\frac{1}{4}\xi^T \gamma \xi + i d^T \xi \right]
\end{equation}
where $d$ is the displacement vector (with  $d_j=\tr(\rho \hat{R}_j)$) and
$\gamma$ is the covariance matrix with entries $\gamma_{jk}=2\tr[\rho
(\hat{R}_j-d_j)(\hat{R}_k-d_k)]-iJ^{(n)}_{jk}$ that completely define the Gaussian state.  $J^{(n)}$ is the symplectic
form for the multimode system:
\begin{equation}
J^{(n)}= \bigoplus_{i=1}^{n}J, \; \; J=\left(
\begin{array}{cc}
0&1\\
-1&0
\end{array}
\right) .
\end{equation}

Gaussian channels are defined as mappings that transform
gaussian states into gaussian states. They can be expressed as
transformations of $\gamma$ and $d$:
\begin{eqnarray}
\gamma&\mapsto& X\gamma X^T + Y\\
d&\mapsto& X d
\end{eqnarray}
Complete positivity of the channel is guaranteed by the condition:
\begin{equation}
Y+i J - i X^T J X \geq 0.
\label{gaussian_channel_cnd}
\end{equation}

We now show how to determine $X, Y$ for an arbitrary gaussian channel
$\Phi$. Recall that the action of any general channel is given by:
$\Phi\left(\rho_{s}\right)=\tr_{e}\left[\hat{U}(\rho_s\otimes\rho_e)\hat{U}^\dagger\right]$,
where $\hat{U}=\exp(-\mathrm{i}\hat{H})$ is a unitary operation
generated by a Hamiltonian $\hat{H}$.  Gaussian channels are generated
by Hamiltonians $\hat{H}$ that are quadratic in the canonical operators:
$\hat{H}=\mathrm{i} \hat{R}^T h \hat{R}$, were $h$ is $2n\times2n$ hermitian matrix
\cite{gaussChannBig}.  Here $\rho_s$ is the input state and $\rho_e$
is the state of environment. Now, for gaussian channels, both $\rho_s$
and $\rho_e$ are gaussian states with covariance matrices $\gamma_s$,
$\gamma_e$ and displacement vectors $d_s$, $d_e$ respectively. The
displacement of the output state depends linearly on $d_e$. As any
displacement of output states by a constant vector is a unitary
operation and as such it does not influence the channel capacity, we assume
that $d_e=0$. The action of $\hat{U}$ on the canonical observables can
be identified with the linear transformation $\hat{U}^\dagger
\hat{R}^T \hat{U}= M\hat{R}^T$. Now, we express $M$ in  block form
with respect to a system/environment partition:
$M= \left(
\begin{array}{cc}
M_{ss}&M_{se}\\
M_{es}&M_{ee}
\end{array}
\right)$.
From teh latter one  obtains  $X=M_{ss}$ and $Y=M_{se} \gamma_{e}M^T_{se}$.

Finally, note that in the context of quantum gaussian channels, power
constraints are usually expressed as a limitation on a mean number of
photons transmitted per channel use.

Squeezed states represent an important class of gaussian states for
communication tasks.  A one--mode squeezed state saturates the
Heisenberg uncertainty principle, with lower quantum noise (variance)
in one of the quadratures as compared with a coherent state. In the
photon number basis a one mode vacuum squeezed state has the
following form:
\begin{equation}
\ket{\zeta;0}	= \sqrt{ \text{sech}\; r}\sum_{n=0}^\infty \frac{\sqrt{(2n)!}}{n!}\left[-\frac{1}{2}e^{i\phi}\tanh r \right]^n \ket{2n}
\end{equation}
where $r$ is the squeezing parameter. In terms of the covariance
matrix formalism, the $\phi = 0$ squeezed vacuum state is described by
\begin{equation}
\gamma =
\left(
\begin{array}{cc}
e^{-2r}&0\\
0&e^{2r}
\end{array}
\right)
\end{equation}
and displacement vector $d=0$.  Displacing a squeezed vacuum state
using the displacement operator $D_{\tilde{d}}$ leads to a state with
unchanged covariance matrix but with the displacement vector
$d=\tilde{d}$. In the two-mode case,  we shall utilize  the two-mode
squeezed vacuum state, with squeezing of the relative position
$x_1-x_2$ and
total momentum  $p_1+p_2$.  The covariance
matrix of this state takes the form \cite{gaussChannBig}:
\begin{equation}
\gamma = H^T diag( e^{-2r},e^{2r},e^{2r},e^{-2r})H
\end{equation}
where:
\begin{equation}
H=\frac{1}{\sqrt{2}}\left(
\begin{array}{cccc}
1&0&-1&0\\
1&0&1&0\\
0&1&0&-1\\
0&1&0&1
\end{array}
\right)
\end{equation}
while the displacement vector $d=0$.

Lastly, for calculation of channel capacities we shall require the
entropy of $n$-mode gaussian states $\rho $. This is given by the
formula \cite{gaussChannBig}:
\begin{equation}
S(\rho)=\sum_{j=1}^n g(\frac{v_j-1}{2})
\end{equation}
in terms of normal modes of the system.  Here $g(x) =
(x+1)\ln(x+1)-x\ln x$ is the entropy of a normal mode with average
occupation number $x$. The $v_j$'s are the symplectic eigenvalues of
the covariance matrix $\gamma$ corresponding to the state $\rho $, {\it
  i.e.}  the square roots of the eigenvalues of the matrix $-J^{(n)}
\gamma J^{(n)} \gamma$. (Note that the symplectic spectrum for each mode is
doubly degenerate and that in the entropy formula each value is taken only
once).

\subsection{Locality rule for classical gaussian MAC}
\label{locality_rule}

The analysis of capacity regions is more intricate in the CV case than
in the discrete variable case, already for classical channels. This is
intimately related to the fact that the capacities are dependent on
power constraints which may lead to various scenarios.  To see this,
consider an example of two 1-to-1 classical channels channels $\Phi_1$
and $\Phi_2$ with noise levels $N_1$ and $N_2$ and the same power
constraints $\tilde{P}$.  We assume that $N_1<N_2$. Suppose each
channel works separately, then $\Phi_i$ achieves the capacity
$\tilde{C}_i=\frac{1}{2}\log\left[1+\tilde{P}/N_i\right]$
\cite{CoverThomas}. Now suppose the channels work in a parallel
setup. The sender aims to maximize the total capacity
$C_T=C_1+C_2=\frac{1}{2}\left(\log\left[1+P_1/N_1\right]+\log\left[1+P_2/N_2\right]\right)$
where $P_i$ is the power allocated to channel $\Phi_i$.  One demands
that the total power available to the user in this case is identical
to the total power used when the channels were utilized separately,
{\it i.e.}  $P_1$ and $P_2$ obey the constraint $P_1+P_2\leq
2\tilde{P}$.  Now since the noise levels $N_1, N_2$ are different, the
senders can increase the total capacity by allocating more power in
the transmission through the channel with the lower level of
noise. When $N_1+2\tilde{P}<N_2$, the optimal choice is to put
$P_1=2\tilde{P}$. In the other case, the optimal allocation is
determined from the relation $N_1+P_1=N_2+P_2$. Using this power
redistribution, the sender can achieve capacity $\tilde{C}> C_1+C_2$.
This process of optimisation is the so--called waterfilling scheme(see {\it e.g.}
\cite{CoverThomas}).

Thus, we see that for Gasussian channels the additivity theorem of
Sec.\ref{discrete_additivity_theorem} cannot be stated as such.
However, observe that the local rates depend only on the local power
constraints (cf. Eq.~(\ref{classcapreg})).
This means that, in a multiuser scenario, adding a resource (channel
or energy) to one sender never helps the others beat their maximal
achievable rates (power constraints pertaining to different users are
not allowed to be combined, hence no inter-user waterfilling effect
can take place).  We call this observation the locality rule for classical
Gaussian MAC's \cite{gaussian_mac} and shall treat it as the appropriate
analog of the additivity rule for classical capacities for discrete
channels.


\subsection{Comparison of strategies which lead to increasing
  transmission rates in case of gaussian state encoding and homodyne
  detection.}
\label{strategies_comparison}

\begin{figure}
	\centering
		\includegraphics[scale=0.5]{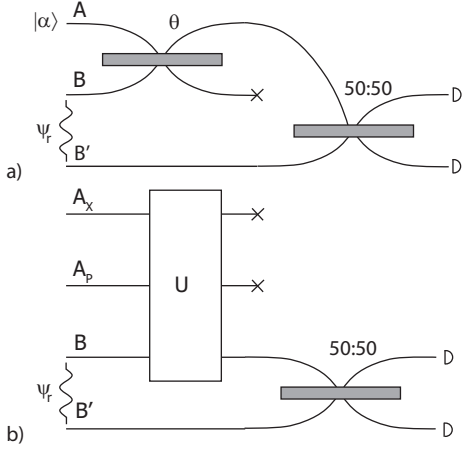}
        \caption{Gaussian channels exhibiting violation of locality
          rule on individual transmission rates: a) beam splitter
          channel $\Phi_\theta$; b) triple QND sum gate channel
          $\Phi$.  The channels were first presented in
          Ref.~\cite{gaussian_mac}.
	\label{fig:gausschannels}}
\end{figure}

We will first study the setup presented in
FIG. \ref{fig:gausschannels}(a). It consists of the channel
$\Phi_\theta$, which is an asymmetrical beam-spliiter with two one-mode input lines and a one mode output. The channel $\Phi_\theta$ works in parallel with a one mode
ideal (identity) channel $\idchannel$. Sender $A$ has access to the input $A$ of
channel $\Phi_\theta$ and sender $B$ to the input $B$ of channel
$\Phi_\theta$ and $B'$ of channel $\idchannel$. The signals from the input
lines $A$ and $B$ are mixed in the channel $\Phi_\theta$ on the beam
splitter, which has transitivity $T=\cos^2 \theta$. The receiver has access to only one output mode of the beam splitter, the second being blocked. The channel $\Phi_\theta$ is thus characterized by a loss $\cos^2\theta
N_A$ of the power input $N_A$ through line $A$.

We place the following power constraints, expressed in terms of mean number of photons used by senders:
\begin{eqnarray}
\text{\{sender } A \text{ av. \# photons \}} &=& N_A\\
\text{\{sender } B \text{ av. \# photons \}} &=& N_B
\end{eqnarray}

We now compare how the choice of quantum states used in the
communication protocol influences the transmission rate
$R_A(\Phi_\theta\otimes\idchannel)$ for sender $A$, while sender $B$
is a helper-sender all the time and its rate
$R_B(\Phi_\theta\otimes\idchannel)=0$. We will also point out
cases where the locality rule is broken. We consider only transmission of
Gaussian states.
We shall focus our attention on the following protocols:

\begin{enumerate}
\item \label{coherentstatestransm} Senders $A$ and $B$ transmit
  coherent states. $A$ encodes messages as displacements of both
  canonical variables of the vacuum state while the probability of
  displacement is chosen, as is standard, to be a Gaussian distribution $p(x,p)=\frac{1}{2 \pi\sigma^2}
  \exp(-\frac{x^2+p^2}{2\sigma^2})$ with $\sigma^2=2N_1$.  Sender $B$
  transmits a fixed chosen coherent state all the time.  The receiver performs
  homodyne detection on both quadratures to  decode the message. This is a
  typical setup for transmission of information through optical
  fibers \cite{optical_networks_practical_perspective}. The achievable rate depends only on the output power corresponding to user $A$ and reads:
\begin{equation}
R_A^{coherent}\leq \log (1 + \sin^2 \theta N_A)
\end{equation}
This rate refers to the case of a lossy channel with transmitivity
$T=\sin^2 \theta$ 
in case when sender performs encoding in coherent
states and receiver performs homodyne detection on the output \cite{caves_quant_limits}.  It
depends only on power constraints for $A$ and manifestly obeys the locality rule.
	
\item Senders $A$ and $B$ use single-mode squeezed vacuum states. Both users
  transmit states which are squeezed in the same canonical variable,
  say $x$. $A$ encodes his message in the displacement of the variable $x$ of his
  state, whose value is Gaussian distributed
  with variance $\sigma_x^2$. The receiver performs homodyne detection only on $x$.
  This setup was studied in \cite{yen_shapiro_mac_sq}. The rate is given
  by:
\begin{equation}
R_A^{squeezed}=\frac{1}{2}\log\left[1+ \frac{\sigma_x^2 \sin^2 \theta}{\sin^2 \theta\! e^{-2R} + \cos^2 \theta\! e^{-2r}}\right],\label{eq:onemodesqcap}
\end{equation}
 where $R$ and $r$ denote the
squeezing parameters of the $x$ quadrature  for senders $A$ and $B$
respectively. The energy constraints can be written as: $\sigma_x^2 \leq 4(
N_A - \sinh^2 R), \sinh^2 r \leq N_B$. User $A$ performs optimization
of the parameter $R$, that is he optimizes the power allocation between
squeezing and mean square displacement. For fixed $\theta$, in the limit
$N_A\rightarrow\infty,N_B\rightarrow\infty$ we get asymptotically
\begin{equation}
R_A^{squeezed} = \log\left[ 1 + N_A\right]\label{eq:onemodesqasymp}.
\end{equation}


	\item \label{entencodingbssect}
	
      Sender $A$ again sends coherent states, encoding his message
      in the displacement of both canonical variables. The
      displacement has probability density distribution as in case
      (\ref{coherentstatestransm}). Sender $B$ transmits a two--mode
      squeezed state,  one mode  through $\Phi_\theta$ and the
      second one through the extra resource $\idchannel$.  The receiver has access to the
      output of $\Phi_\theta$ and $\idchannel$. The decoding consists of a joint
      measurement of the canonical variables $x_{\Phi_\theta}-x_I$ and
      $p_{\Phi_\theta}+p_I$ on the output modes of the setup. To achieve this, the output modes of $\Phi_\theta$ and $\idchannel$
   are mixed  on a $50:50$ beam splitter followed by homodyne measurements of $x_1$ and $p_2$
      on the output modes of the $50:50$ beam splitter.  In this
      setup, the sender $B$ is assumed to make use of entangled states.

The formula for the transmission rate is now given by:
\begin{equation}
R_A^{ent}=\log\left[1+\frac{\sigma^2 \sin^2\theta}{2(\cosh r - \cos\theta\sinh r)^2}\right].
\label{entcap}
\end{equation}
Here $\sigma^2 = 2 N_A$ is the variance in the displacement of
canonical variables in $A$'s mode. Here, $r$ denotes the squeezing
parameter of the two mode squeezed state sent by $B$. The imposed power
constraints imply that $\sinh^2 r = N_B/2$. For given $N_A$, the optimal values of
(\ref{entcap}) lie on curve:
\begin{equation}
\cos\theta = \tanh r \label{eq:squeeztheta}
\end{equation}
which leads to the following  maximal rate formula:
\begin{equation}
R^{ent-opt}_A=\log\left[1+N_A\right].
\end{equation}
$R_A^{ent-opt}$ is in fact equal to the rate achievable
by a one mode ideal channel case when sender performs encoding in
coherent states and receiver performs homodyne detection on the
output. Thus entanglement can be used to completely overcome power
loss in the case of coherent state encoding. The same effect is also obtained in the second case, without entanglement, as described above, but only in the asymptotic regime of infinite power (see Eq.(\ref{eq:onemodesqasymp}). 

It is indeed worth noting, for comparison, that the limit $N_A\rightarrow\infty,N_B\rightarrow\infty$ of Eq.(~\ref{eq:onemodesqcap}) under the constraint Eq.(~\ref{eq:squeeztheta})
leads to:
\begin{equation}
R_A^{squeezed} \leq \frac{1}{2}\log\left[1+16 N_A\right]\approx\frac{1}{2}\log\left[1+ N_A\right] =\frac{1}{2} R^{ent-opt}_A.
\end{equation}
Comparison of this result with Eq.(~\ref{eq:onemodesqasymp}) shows that one mode squeezed states transmission requires much higher squeezing  to reach the rates achievable by two mode squeezed state transmission.

\end{enumerate}	

We can also calculate two upper bounds $R_A(\Phi_\theta)$ for
transmition rates only through channel $\Phi_\theta$:
\begin{enumerate}
\item Bound based on maximal entropy of a state with mean number of
  photons equal to the mean number of photons in the output mode of
  the channel $\Phi_\theta$. We shall refer to it as to the {\it
    output entropy bound}.  This tells us how large a rate is
  achievable if no entanglement is allowed in the communication
  protocol and is given by:	
\begin{eqnarray}
R_A^{prod-bound}&=&g(N_{out})\\
&=&g\left(\sqrt{N_A}\sin^2\theta+\sqrt{N_B}\cos^2\theta\right)
\label{upper_bound_bs}
\end{eqnarray}	

\item Bound based on maximal entropy of a state with mean number of
  photons equal to the mean number of photons in the input mode $A$ of the
  channel $\Phi_\theta$.  This may be referred to as an {\it input
    entropy bound}. This bound cannot be violated by any type of
  communication protocol, entanglement-free or entanglement-aided, and
  it tells us how much information can be transmitted with given
  energy constraints if sender $A$ is connected to the receiver by a
  one mode ideal line. We will check how close the protocols described
  above approach this bound, which is given by
	\begin{equation}
	R_A^{max}\leq g(N_A).
	\end{equation}
	\end{enumerate}
The bounds $R_A^{prod-bound}$ and $R_A^{max}$ allow us to express the theoretical maximum
rate for sender $A$ in the form
$\min(R_A^{prod-bound},R_A^{max})$.


\begin{figure}
	\centering
		\includegraphics[scale=0.6]{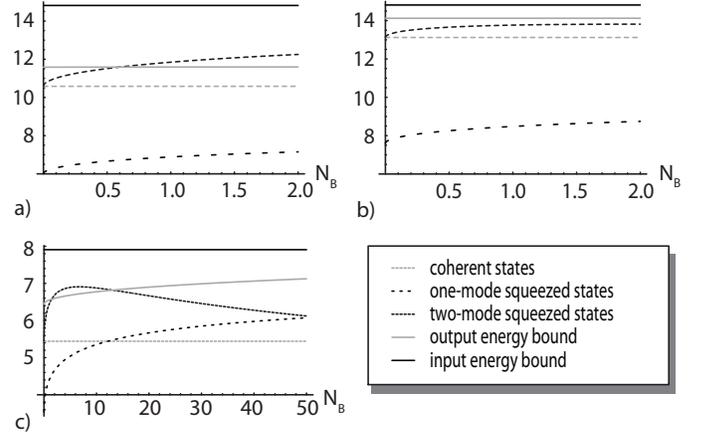}	
	\caption{
	Rates which are achievable by different encoding  schemes for the beam splitter channel $\Phi_\theta$ as a function of the energy constraint $N_B$ for sender $B$. Rates are evaluated for the following setups of channel parameters $\theta$ and energy constraints $N_A$ for sender $A$;
	A) $\theta = \pi/4,N_A = 10^6$ ;B) $\theta = 0.2,N_A = 10^6$; C) $\theta = 0.5,N_A = 10^3$}
	\label{fig:squeezing1}
\end{figure}

FIG. \ref{fig:squeezing1} shows the behaviour of rates achievable by
different encoding schemes and parameter regimes as a function of the energy constraint $N_B$ for
sender $B$ (the schemes and bound correspond directly to the points
(1-5) in the main text). Figure a) presents the situation where using
entangled states quickly becomes more efficient than using any product state encoding, while on the other hand  Fig. b) presents
a situation where entanglement cannot beat the upper bound for rates
achievable for product states encoding.
In Fig. c) we consider the behaviour of rates for large range of values of $N_B$. We can observe
that in the low $N_B$ range, the strategy using entanglement states is the best among the
three considered approaches. However increasing $N_B$ leads to a maximal value of
$R^{ent}_A$ after which further growth leads to a diminishing rate. This can be explained by increasing of the
entanglement of the output state with the erased mode. In case of
$R_A^{squeezed}$ the situation looks different. It was shown in
the paper \cite{yen_shapiro_mac_sq} that in the limit
$N_A\rightarrow\infty,N_B\rightarrow\infty$ rate $R^{squeezed}_A$
asymptotically approaches the upper bound for the transmission rate for sender
$A$ expressed by $R_{A}^ {max}$.  It has to be reiterated here that this
strategy requires extremely high squeezing to approach the maximal rate
achieved by the protocol using a two mode squeezing scenario.

\begin{figure}
	\centering
		\includegraphics[scale=0.6]{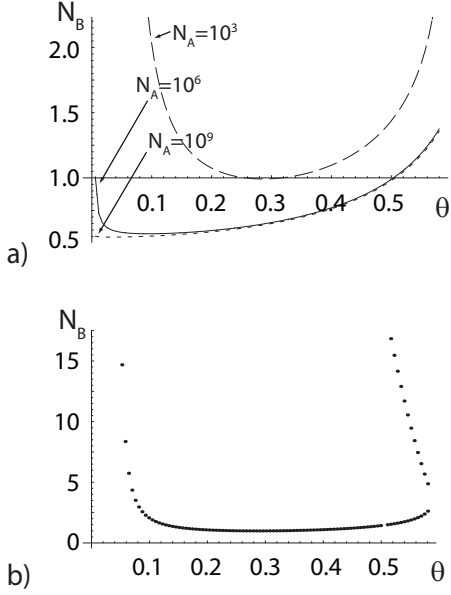}
        \caption{ a) Lines present bounds of the areas where
          $R_A^{ent}>R_A^{prod-bound}$. They are given by the condition
          $R_A^{ent}/R_A^{prod-bound}=1$. Lines refer to the cases of
          $N_A=10^3,10^6,10^9$. For given $N_A$ the superadditive region
          lies above the corresponding line. b) The dotted line delimits the area
          where $R_A^{ent}>R_A^{prod-bound}$ for $N_A=10^3$. Notice that the superadditive region is considered in a large scale of $N_B$. The figure shows that for there is large window of parameters
          $N_B,\theta$ for which $R_A^{ent}>R_A^{prod-bound}$ can be
          observed.}
	\label{fig:supaddstart}
	
\end{figure}

In case of the protocol using two mode squeezed states it is interesting
to ask about the lower limit of squeezing for which the rate achieved  starts to be higher than the rate achieved by any protocol
based only on product states encoding.
FIG. \ref{fig:supaddstart}.a) presents demarcation curves
$R_A^{ent}/R_A^{prod-bound}=1$ in  the $\theta-N_B$ parameter plane for
three different values of the parameter $N_A$.  For fixed $N_A$, with
increasing $N_B$, we move above the demarcation curve and fall into the area
where $R_A^{ent}>R_A^{prod-bound}$.  The minimal mean photon number in the
entangled state, required to approach this area, amounts to around
$N_B=1,0.6,0.55$ for $N_A=10^3,10^6,10^9$. These values of $N_B$
refer to the following squeezing levels which are experimentally realistic: $5.72 dB, 4.55 dB, 4.37 dB$. The demarcation
curve is crossed as $\theta$ equals $0.28,0.1,0.02$ or
transmitivity $0.077,0.01,0.0004$.
For large $N_A$, the locality rule is broken for $\theta\approx0$. In this regime the setup reproduces the continuous variable dense coding scheme.
 $N_B=1$ means that we use two mode squeezed state with squeezing
 equal $5.72$dB which is reasonable value for experimental setup. In
 FIG. \ref{fig:supaddstart}.b). we change the scale of observation and show that
 breaking of the locality rule occurs for quite a large range of the parameter
 $N_B$ and $\theta$.

\subsection{Realization of XP gate by linear optics and one mode squeezed states. Influence of noise on superadditivity effect}
\label{noise_and_realisation}

In this section, we shall start with  details of realization of the three
input quantum non demolition channel $\Phi$ presented in FIG. \ref{fig:gausschannels}(b).
This will lead naturally to a discussion of the interplay between and noise (or imperfections) and superadditivity.

The channel $\Phi:A_XA_P B \mapsto R$ acts as follows
$\Phi(\rho_{A_XA_P}\otimes\rho_{B})=\tr_{A_XA_P}\left[\hat{U}(\rho_{A_XA_P}\otimes\rho_{B})\hat{U}^\dagger\right]$. Sender
$A$ holds lines $A_X$ and $A_P$, while sender $B$ holds line
$B$. $\hat{U}$ is an unitary operator of the  form $\hat{U}=\exp [-i
(\hat{x}_X\hat{p}_B-\hat{p}_P\hat{x}_B)]$, which can be factorized as follows:
\begin{eqnarray}
\hat{U}&=&\exp [-i (\hat{x}_X\hat{p}_B-\hat{p}_P\hat{x}_B)] \\
&=&\exp[\frac{i}{2} \hat{x}_X\hat{p}_P] \exp[-i \hat{x}_X\hat{p}_B]\exp[i \hat{p}_P\hat{x}_B].
\label{xp_u}
\end{eqnarray}
The $XP$ interaction, appearing here,  can be obtained by measurement-induced
continuous-variable quantum interactions as described in
Ref.~\cite{xpcv_backround}. An experimental proof--of--concept has been presented in  Ref.~\cite{xpcv_realization}.

The superadditive effect of single user capacity (breaking of the locality rule) for this channel was considered by us in \cite{gaussian_mac} for this channel, where the $XP$ interaction was assumed to be ideally implemented.
However,  the method of measurement-induced continuous-variable quantum interactions
is, in practice, imperfect and introduces errors in the output states.  To study such
errors it is useful to write down how canonical observables are
transformed by the realization of the $XP$ gate \cite{xpcv_backround,xpcv_realization}:
\begin{eqnarray}
\hat{x}^{out}_1&=&\hat{x}^{in}_1-\sqrt{\alpha}\hat{x}_0-\sqrt{\beta}\hat{x}_{S_1},\\
\hat{p}^{out}_1&=&\hat{p}^{in}_1-\frac{1-T}{\sqrt{T}}
\hat{p}^{in}_2+\sqrt{\alpha/T}\hat{p}_0+\sqrt{T\beta}\hat{p}_{S_2},\label{eq:cvtransf2}\\
\hat{x}^{out}_2&=&\hat{x}^{in}_2+\frac{1-T}{\sqrt{T}}\hat{x}^{in}_1-\sqrt{\alpha/T}\hat{p}_0+\sqrt{T\beta}\hat{x}_{S_1},\label{eq:cvtransf3}\\
\hat{p}^{out}_2&=&\hat{p}^{in}_2-\sqrt{\alpha}\hat{p}_0+\sqrt{\beta}\hat{p}_{S_2}
\end{eqnarray}
where: $\alpha=(1-T)(1-\eta)/(1+T)\eta,\beta=(1-T)/(1+T)$,
 $\hat{x}_{S_1},\hat{p}_{S_2}$ are canonical
observables of two different modes in squeezed states, $\eta$ is efficiency
of the homodyne detectors inside the $XP$ gate realisation
and $\hat{x}_0,\hat{p}_0$ are canonical observables of two different modes in
the coherent states used that homodyne detectors.
Paramter $T$ depends on the configuration of the $XP$ gate realisation and can be manipulated.
Choosing
$T=\frac{1}{2}\left(3-\sqrt{5}\right)$, the coefficients of
$\hat{p}^{in}_2$ and $\hat{x}^{in}_1$ in
Eqs.(~\ref{eq:cvtransf2})-(\ref{eq:cvtransf3}) become $-1$ and $1$.  We
will hereby represent this XP gate realization as a quantum noisy channel described by
transformation matrices $X,Y$, using the method outlined in Sec. \ref{cv_section}. Here we assume that errors introduced by linear optical elements can be neglected  in comparison with natural noise due
to the physical generation of highly squeezed states. Then
\begin{eqnarray}
X=\left(
\begin{array}{cccc}
1&0&0&0\\
0&1&0&-1\\
1&0&1&0\\
0&0&0&1
\end{array}
\right),\!
Y=\left(
\begin{array}{cccc}
\sigma_1^2&0&0&0\\
0&\sigma_2^2&0&0\\
0&0&\sigma_2^2&0\\
0&0&0&\sigma_1^2
\end{array}
\right)
\end{eqnarray}
with $\sigma_1^2=\alpha +\beta e^{-2 s},\sigma_2^2=\alpha/T +\beta T e^{-2 s}$. We assumed that  squeezed states in both modes  have the same squeezing level.
This gate reproduces the ideal XP gate in the limit of infinite squeezing $s\rightarrow\infty$ and ideal homodyne detection $\eta\rightarrow 1$.
If all $XP$ gates used in the implementation of the considered channel
have the same parameters, we can collect all noise components under the common factor $\sigma_{noise}^2=\sigma_1^2+\sigma_2^2$.


Now suppose that sender $B$ also has access to the input $B'$ of a one mode ideal channel $\idchannel$ (see Fig. \ref{fig:gausschannels}(b)). We are interested in the maximal rate $R_{A}^{(1)}(\Phi\otimes\idchannel)$ for  sender $A$ under the following protocol  \cite{gaussian_mac} when sender $A$ transmits displaced squeezed one mode vacuum states. States transmitted through line $A_X$ ($A_P$) are squeezed in the canonical observable $\hat{x}$ ($\hat{p}$),  where the squeezing parameter is $R$. Sender $A$ encodes his message in the displacement of canonical observables $\hat{x}$ ($\hat{p}$) for line $A_X$ ($A_P$). As usual, the displacement has Gaussian distribution with variance $\sigma^2$.
Sender $B$ continuously transmits a constant fixed two mode squeezed vacuum state, with squeezing parameter $r$. One mode is transmitted through line $B$ and the other through  line $B'$. The receiver performs joint homodyne detection of $x_B-x_{B'}$ and $p_B+p_{B'}$ on the output of channels $\Phi$ and $\idchannel$ to decode the message. The rate in this case is now calculated to be:
\begin{eqnarray}
R_A^{(1)}=\log\left(1+\frac{\sigma^2}{e^{-2R}+\frac{e^{-2r}}{2}+\frac{\sigma_{noise}^2}{2}}\right) \label{eq:phinoisecap}
\label{noise_1}
\end{eqnarray}
The imperfections in the implementation of the desired unitary evolution appear in the form of the extra noise term $\sigma_{noise}^2/2$ in the expression for maximal rate, as compared to the ideal case described in \cite{gaussian_mac}.

For a more realistic description, we shall model the influence of various unavoidable imperfections, associated {\it e.g.} with the  implementation of displacement during encoding, the measurement process realized by the receiver for decoding, and the interaction with environment at finite temperature, by  thermal noise channels.
This type of a channel is a 1-to-1 lossy channel mixing an input state with a thermal state $\rho_{N_{Th}}$ containing an average of $N_T$ photons, at a beam splitter with transmissivity $T=\cos^2 \omega$. The receiver receives only one of the output modes of this mixing beam splitter. The covariance matrix $\gamma$ of an input state is transformed by this channel as follows
\begin{eqnarray}
\gamma\mapsto T\gamma+ (1-T)\gamma_{N_{Th}},
\end{eqnarray}
where $\gamma_{N_{Th}}=N_{Th}\mathbb{I}$.
Below we assume that $XP$ gate is perfect, however we put the thermal noise channels parameterized by $\omega$ and $N_{Th}$ at two places: between the output of the non-demolition channel $\Phi$ and receiver and between the output of supporting channel $\idchannel$ and receiver.
Now the transmission rate for the upper sender, using the same protocol as described earlier in the noiseless case, is modified and is calculated here to be
\begin{eqnarray}
R_{A}^{(1)}\!=\!\log\!\left(\!1\!+\!\frac{\sigma^2 \cos^2 \omega}{(e^{-2 R}+\frac{e^{-2r}}{2})\cos^2\omega+(1+N_{Th})\sin^2\omega}\!\right).
\label{noise_2}
\end{eqnarray}

\begin{figure}
	\centering
		\includegraphics[scale=0.5]{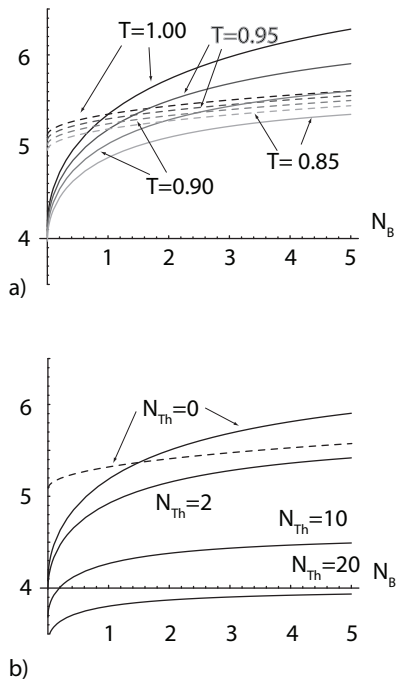}
	\caption{Influence of thermal noise on capacity $R^{(1)}_{A}(\Phi_\theta\otimes\idchannel)$ of the BS channel $\Phi_\theta$ working with parameters $\theta=0.25, N_A=10^3$: a) dependence on transitivity $T$ of thermal noise channel, here $N_{Th}=0$; rates $R^{(1)}_{A}(\Phi_\theta\otimes\idchannel)$ (solid) are compared with output entropy bound $R_A^{prod-bound}$ (dashed) b) dependency on mean photon number $N_{Th}$ in thermal state for $N_{Th}=0,2,10,20$ and an exemplary value $T=0.95$; here we provide output entropy bound $R_A^{prod-bound}$ only for $N_{Th}=0$ since additional thermal noise increases output entropy and makes bound less tight.}
\label{fig:transmchanges}		
\end{figure}

In a similar way, we now also model effects of noise on the beamsplitter MAC channel  $\Phi_\theta$  discussed earlier (Fig. \ref{fig:gausschannels}(a)) in  point \ref{entencodingbssect} of Sec.~\ref{strategies_comparison}. We again place the thermal noise channel (parametrised by $\omega,N_{Th}$) between output of the $\Phi_\theta$ channel and receiver and between output of the supporting channel $\idchannel$
and receiver. In this case, we obtain now the following rate of the upper sender:
\begin{equation}
R_{A_1}\!=\!\log\!\left(\!1+\!\frac{\sigma^2 \sin^2 \theta T}
{\!(\cosh\! r \!- \cos\! \theta \sinh\! r)^2T\! +\! (1+\!2 \!N_{Th})(1\!-\!T)\!}\right).
\label{noise_3}
\end{equation}
Here $\theta$ is the parameter of the BS channel $\Phi_\theta$, $T = \cos^2 \omega$ is the transmissivity of the thermal noise channel and $N_{Th}$ is the mean photon number of the environment.
In Fig. \ref{fig:transmchanges} we use this result to illustrate how the capacity $R_{A}^{(1)}(\Phi_\theta\otimes\idchannel)$ changes with the parameters of the thermal noise channel. Even if the effect of thermal noise channel is small - its transmitivity is large and   $N_{Th}=0$ - the capacity gain becomes negligible and from $T=0.85$  no enhancement over the upper bound for rates obtained using product codes is observed. This scenario corresponds to the case where there are only losses in the thermal channel.

Now we use the presented results to discuss the possibility of experimental verification of the considered superadditive effects in the context of available technological resources \cite{hv,iqoe}.
For  homodyne detection we will assume quantum efficiency at the level $\eta=99\%$ as in\cite{qeffexp} and the dark noise level at 20dB below the shot noise of the local oscillator. For power constraints of sender $A$: $N_A=1000$ we
remain in the regime of linear approximation of homodyne detection (mean number of photons of local oscillator is on the level $4\times 10^6$). Note finally, that the highest observed value of single mode squeezing \cite{hv} is at the level of $10$ dB which corresponds to mean number of photon $2.025$. 

We start with a discussion of the setup $\Phi\otimes\idchannel$ in the context of the implementation of the $XP$ gate presented in \cite{xpcv_realization}. In that experiment,  squeezing of $5.6$ dB and detectors with quantum efficiency $\eta=98\%$ were used. A quadrature transfer coefficient of $T_P=SNR_{out}/SNR_{in}=0.4$ was reported, where $SNR_{in (out)}$ is the signal-to-noise ratio for the signal input (output) of the gate. This coefficient $T_P$ can be translated to $\sigma_{noise}^2=5$ in the model of the $\Phi$ gate. Assuming this value in Eq.(~\ref{eq:phinoisecap}) and comparing the rate with the output entropy bound leads to the conclusion that  the superadditivity effect described here cannot be achieved for  these parameters.  On the other hand,  implementation of the $XP$ gate with the use of $10$ dB squeezing leads to $\sigma^2_{noise}=0.098$ and moves the transmission into the supperadditivity effect regime.

For the setup $\Phi_\theta\otimes\idchannel$ the situation is more optimistic.
With a  realistic loss level of $5\%$ on the optical elements and homodyne detection efficiency as described above one gets $\cos^2\omega = 0.94, N_T=0.09$. Our results show that the supeadditivity effect can be observed for $\theta=0.25$ for squeezing upwards of  $7.8$ dB (mean photon number $2.1$). Thus one can draw the conclusion that a loop-hole free verification of the superadditivity effect can be done with the present state of art  quantum optical experimental techniques.

Formulas (\ref{noise_1}, \ref{noise_2} and \ref{noise_3}) can be understood in a generic signal-to-noise "fenomenological" scheme as $R_A = \log(1+\sigma^2_{signal}/\sigma^2_{noise})$. The variance $\sigma^2_{signal}$ describes how spread out are the input states of the sender  in phase space and $\sigma^2_{noise}$ describes the effective noise level associated with measurement of displacement which here is the carrier of classical information. Senders can manipulate  $\sigma^2_{signal}$ and $\sigma^2_{noise}$ by changing energy allocation used for displacement and squeezing. In this way one can bring nearer to the bound for channel capacity.  Noise introduced by imperfection of elements of the communication setup  plays the role of a lower bound for $\sigma^2_{noise}$ and the user can not decrease measurement error below its level.

\section{Conclusions}
\label{conclusions}

The superadditivity of classical capacity regions have been reported previously in
case of discrete \cite{discrete_mac} and continuous variables (Gaussian) \cite{gaussian_mac} cases.
Here we have analysed these problems in more detail. We have been able to show that asymmetry
of the channel is not crucial for the superdditivity effect. Even more interestingly, we have proven
explicitly that two-input entanglement is not enough in some cases - in our case we have shown analytically
that at least 5-input entanglement is needed. It is interesting that here the
5-qubit code error correction code has been used (see \cite{chuang_nielsen}) to beat the C-type (classical) multi-access capacity
which so far was a tool related to Q-type (quantum) bipartite capacity.  Moreover
we do not know of any example in bipartite classical capacity where
 more than two-input entanglement is needed to achieve the asymptotic bound.
In fact the celebrated effect of breaking of additivity of Holevo function \cite{Hastings}
needs two copies of the channel. We believe that our result will inspire the search for
requirement of multipartite entanglement for achieving the asymptotic Holevo capacity
in the bipartite case.
In both bipartite and multiuser cases, this opens the intriguing question concerning
which types of multipartite entanglement (bipartite quantum codewords, cluster, Dicke-type etc.)
are the best for  achieving asymptotic  classical capacities. We leave this type of
questions for further research.

In case of continuous variables, we carefully compared different scenarios
including the one described by Yen and Shapiro Ref. \cite{yen_shapiro_mac_sq}. We incorporated explicitly imperfection of the schemes into calculations. The
success of superadditivity depends on power of the light used and may be destroyed
by thermal noise or even by loses if they are large enough.
On the other hand we found that the condition for two mode squeezing used for the effect
may  not be very demanding ($4.55dB$ 
). This opens the possibility of experimental
confirmation of the effect in near future.
Again the question of  channels that require multipartite CV-type entanglement
is quite natural and may be not easy in case of Gaussian channels.

\section{Appendix}

Here we prove that capacity regions of discrete classical $n$-to-$1$ channels are additive:
\begin{equation}
\mathcal{R}(\Phi_I\otimes\Phi_{II})=\mathcal{R}(\Phi_I)+\mathcal{R}(\Phi_{II}),
\end{equation}
where $\mathcal{R}(\Phi_I)+\mathcal{R}(\Phi_{II})=\{u_I+u_{II}:u_I\in\mathcal{R}(\Phi_I),u_{II}\in\mathcal{R}(\Phi_{II})\}$.
For simplicity it is assumed that both channels, $\Phi_I, \Phi_{II}$ have the same number of senders. This situation is easy to obtain by formal extension of the set of senders for one of the channels. The messages transmitted by these additional senders are then always lost.
We will use short-hand notation $R_S=\sum_{i\in S}R_i$ for vector of rates $R\in\mathbb{R}^n$ where $R_i$ is $i$-th element of $R$ and $S\subseteq E$ is a subset of senders $E$.
The proof for the $n$-to-1 case follows the same principles as for 2-to-1 channels. Here we provide only the parts that are distinct from that latter case.

We start with $\tilde{\mathcal{R}}(\Phi_I\otimes\Phi_{II})\subseteq\bar{\mathcal{R}}(\Phi_I\otimes\Phi_{II})\subseteq\mathcal{R}(\Phi_I\otimes\Phi_{II})$.
Let us assume that senders transmit with rates given by vector $\tilde{R}$.
As in 2-to-1 case this vectors has to belong to the fixed probability capacity region for input symbols probability distribution $\tilde{p}=p(Q^I,Q^{II}) \prod_i p(X^I_i,X^{II}_i|Q^I,Q^{II})$.
Below we provide upper bound for this region:
\begin{eqnarray}
  &&\tilde{R}_S\leq I(X_S:Y|X_{S^C},Q)\\
  &&=H(Y|X_{S^C},Q)-H(Y|X_S,X_{S^C},Q)\nonumber \\
  &&=H(Y|X_{S^C},Q) \label{eq:logprodprob}\\
  &&-H(Y^I|X_S^I,X_{S^C}^I,Q^I)-H(Y^{II}|X_S^{II},X_{S^C}^{II},Q^{II})\nonumber\\
  &&\leq H(Y^{I}|X_{S^C}^{I},Q^I)+H(Y^{II}|X_{S^C}^{II},Q^{II})\label{eq:entropsubadd}\\
  &&-H(Y^I|X_S^I,X_{S^C}^I,Q^I)-H(Y^{II}|X_S^{II},X_{S^C}^{II},Q^{II})\nonumber\\
  &&=I(X_{S}^I:Y^I|X_{S^C}^I,Q^I)+I(X_{S}^{II}:Y^{II}|X_{S^C}^{II},Q^{II})\label{eq:nsendebound},
\end{eqnarray}
where  Eq.~(\ref{eq:logprodprob}) is based on the factorisation of conditional probabilities defining the channel action for the product channel, while in Eq.~(\ref{eq:entropsubadd}) we use entropy subadditivity.
On the other hand, evaluation of Eq.~(\ref{eq:fixprobdef}) for input symbols probability distribution $\bar{p}=\tilde{p}_I\tilde{p}_{II} = \left( p(Q^I) \prod_i p(X^I_i|Q^I) \right) \left(p(Q^{II}) \prod_i p(X^{II}_i|Q^{II})\right)$ leads to the region:
\begin{eqnarray}
\bar{\mathcal{R}}&=&\{ R\in\mathbb{R}^n: \forall_{i\in E} R_i\geq 0,\\
&&\forall_{S\subseteq E}\; R_S\leq I(X_{S}^I:Y^I|X_{S^C}^I,Q^I)+\nonumber\\
&&I(X_{S}^{II}:Y^{II}|X_{S^C}^{II},Q^{II})\nonumber\}.
\end{eqnarray}
Combining this result with the bound Eq.~(\ref{eq:nsendebound}) it is easy to see that $\tilde{\mathcal{R}}(\Phi_I\otimes\Phi_{II})\subseteq\bar{\mathcal{R}}(\Phi_I\otimes\Phi_{II})$ holds.


It remains to be shown now  that
\begin{equation}
\bar{\mathcal{R}}(\Phi_I\otimes\Phi_{II})=\tilde{\mathcal{R}}(\Phi_I)+\tilde{\mathcal{R}}(\Phi_{II}) \label{eq:fixedprobcapadd},
\end{equation}
where again $\tilde{\mathcal{R}}(\Phi_I)$ ($\tilde{\mathcal{R}}(\Phi_{II})$) is evaluated for margin probability distribution $\tilde{p}_I$ ($\tilde{p}_{II}$).
The following argument is based on the fact that fixed probability capacity region is a polymatroid \cite{HanlyWhiting}.

\Def{1}{1} Let $E=\{1,\ldots,n\}$ and $f:2^E\mapsto \mathbb{R}_+$ be a set functions (i.e. function that maps subsets of $E$ into $\mathbb{R}_+$. The polyhedron:
\begin{equation}
B(f)=\{x\in \mathbb{R}^n: \forall_{S\subseteq E} x_S\leq f(s), \forall_{i\in E} x_i\geq 0\}
\end{equation}
is a polymatroid if the set function $f$ satisfies: (i) $f(\emptyset)=0$, (ii) $S\subseteq T\Rightarrow f(S)\leq f(T)$, (iii) $f(S)+f(T)\geq f(S\cap T)+f(S\cup T)$.

\Le{1}{1} The fixed probability capacity region (cf. Eq.~(\ref{eq:fixprobdef}))is a polymatroid.

\Pro
Observe that conditional mutual information $I(X_S:Y|X_{S^C},Q)$ plays  the role of a set function $f(S)$, in the above equation. All we have to do now is to check conditions $(i)-(iii)$ defining the polymatroid. By definition if there is no sender, mutual information is equal to $0$, which proves $(i)$.
Now, let us write
\begin{eqnarray}
f(T)&=&I(X_T:Y|X_{T^C},Q)\\
&=&H(Y|X_{T^C},Q)-H(Y|X_{T},X_{T^C},Q)\\
&\geq&H(Y|X_{S^C},Q)-H(Y|X_{S},X_{S^C},Q)\label{eq:entropyreduction}\\
&=&I(X_{S}:Y|X_{S^C},Q)\\
&=&f(S)
\end{eqnarray}
where in Eq.~(\ref{eq:entropyreduction}) we use the fact that additional information reduces entropy ($S\subseteq T\Rightarrow T^C\subseteq T^C$) and $S\cup S^C=T\cup S^C = E$. Since subsets $S, T$ are arbitrary, the condition $(ii)$ is satisfied. Similarly, can check the condition $(iii)$:
\begin{eqnarray}
f(S)+f(T)&=&I(X_S:Y|X_{S^C},Q)+\\
&&I(X_T:Y|X_{T^C},Q)\nonumber\\
&=&H(Y|X_{S^C},Q)-H(Y|X_{S},X_{S^C},Q)+\nonumber\\
&&H(Y|X_{T^C},Q)-H(Y|X_{T},X_{T^C},Q)\nonumber\\
&=&H(Y,X_{S^C},Q)-H(X_{S^C},Q)\nonumber\\
&&-H(Y|X_{S},X_{S^C},Q)+\nonumber\\
&&H(Y,X_{T^C},Q)-H(X_{T^C},Q)\nonumber\\
&&-H(Y|X_{T},X_{T^C},Q)\nonumber\\
&=&H(Y,X_{S^C},Q)+H(Y,X_{T^C},Q)\label{eq:sendindep1}\\
&&-\sum_{i_\in S}H(X_i|Q)-H(Q)\nonumber\\
&&-\sum_{i_\in T}H(X_i|Q)-H(Q)\nonumber\\
&&-2H(Y|X_E,Q)\nonumber\\
&\geq&H(Y,X_{S^C\cup T^C},Q)\label{eq:stronsubadd}\\
&&+H(Y,X_{S^C\cap T^C},Q)\nonumber\\
&&-\sum_{i_\in S}H(X_i|Q)-H(Q)\nonumber\\
&&-\sum_{i_\in T}H(X_i|Q)-H(Q)\nonumber\\
&&-2H(Y|X_E,Q)\nonumber\\
&=&H(Y,X_{S^C\cup T^C},Q)\\
&&+H(Y,X_{S^C\cap T^C},Q)\nonumber\\
&&-\sum_{i_\in S\cap T}H(X_i|Q)-H(Q)\nonumber\\
&&-\sum_{i_\in S\cup T}H(X_i|Q)-H(Q)\nonumber\\
&&-2H(Y|X_E,Q)\nonumber\\
&=&H(Y,X_{S^C\cup T^C},Q)\label{eq:sendindep2}\\
&&+H(Y,X_{S^C\cap T^C},Q)\nonumber\\
&&-H(X_{S\cap T},Q)-H(X_{S\cup T},Q)\nonumber\\
&&-2H(Y|X_E,Q)\nonumber\\
&=&I(X_{S\cap T}:Y|X_{(S\cap T)^C},Q)+\\
&&I(X_{S\cup T}:Y|X_{(S\cup T)^C},Q)\nonumber\\
&=&f(S\cap T)+f(S\cup T).
\end{eqnarray}
where in Eq.~(\ref{eq:sendindep1}), Eq.~(\ref{eq:sendindep2}) we use chanin rule (i.e. $H(A,B,C)=H(A|B,C)+H(B|C)+\ldots+H(C)$ )
together with independency of senders (i.e. $H(A|B)=H(A)$ and
in. Eq.~(\ref{eq:stronsubadd}) we use the strong subadditivity of entropy (i.e.
$H(A,B)+H(A,C)\leq H(A,B,C)+H(A)$).

\proofend

We go back to Eq.~\ref{eq:fixedprobcapadd} and verify that $\bar{\mathcal{R}}(\Phi_I\otimes\Phi_{II})\supseteq\tilde{\mathcal{R}}(\Phi_I)+\tilde{\mathcal{R}}(\Phi_{II})$.
This can be done by a direct coordinate sum, i.e
\begin{equation}
\tilde{R}^{I}_S+\tilde{R}^{II}_S\leq I(X_S^{I}:Y^{I}|X^{I}_{S^C},Q^I) + I(X_S^{II}:Y^{II}|X^{II}_{S^C},Q^{II}).
\end{equation}
where $\tilde{R}^{I}_S\in\tilde{\mathcal{R}}(\Phi_I), \tilde{R}^{II}_S\in\tilde{\mathcal{R}}(\Phi_{II})$ and by definition obey Eq.~(\ref{eq:fixprobdef}).

Finally, we will show $\bar{\mathcal{R}}(\Phi_I\otimes\Phi_{II})\subseteq\tilde{\mathcal{R}}(\Phi_I)+\tilde{\mathcal{R}}(\Phi_{II})$.
Since there is equivalence between the vertex and the half space representation \cite{lectures_on_polytopes} of a convex polyhedron, we only have to show that each vertex $v\in\bar{\mathcal{R}}(\Phi_I\otimes\Phi_{II})$ can be expressed as $v=u+w$ where $u$ and $w$ are suitable vertices of $\tilde{\mathcal{R}}_I$ and $\tilde{\mathcal{R}}_{II}$ respectively.

As we have seen, the fixed probability capacity region is polymatroid. This leads to a key property of the set of its vertices \cite{edmonds}. Let $\pi$ be an ordered choice from the set of senders $E$. For each ordered choice $\pi$, there is a vertex $v$ with entries: $v_{\pi_1}=f(\pi_1)$, $v_{\pi_i}=f(\{\pi_1,\ldots,\pi_i\})-f(\{1,\ldots,\pi_{i-1}\})$ and $\forall_{i\notin \pi}v_i=0$. On the other hand, we  can always find an ordered choice $\pi$ which defines given vertex. It may happen that more than one ordered choice gives the vertex with the same entries.
E.g. in the 2-to-1 case, fixed probability capacity region is given by the vertices:
\begin{eqnarray}
\pi=\emptyset&\!\!:\!\!&\left(\begin{array}{c}
0\\
0
\end{array}\right),\\
\pi=\{1\}&\!\!:\!\!&\left(\begin{array}{c}
I(X^I_1:Y^I|X^I_2,Q^I)+I(X^{II}_1:Y^{II}|X^{II}_2,Q^{II})\\
0
\end{array}\right),\\
\pi=\{2\}&\!\!:\!\!&\left(\begin{array}{c}
0\\
I(X^I_2:Y^{I}|X^I_1,Q^I)+I(X^{II}_2:Y^{II}|X^{II}_1,Q^{II})
\end{array}\right)\\
\pi=\{1,2\}&\!\!:\!\!&\left(\begin{array}{c}
I(X^I_1:Y^I|X^I_2,Q^I)+I(X^{II}_1:Y^{II}|X^{II}_2,Q^{II})\\
I(X^{I}_2:Y^{I}|Q^I)+I(X^{II}_2:Y^{II}|Q^{II})
\end{array}\right),\\
\pi=\{2,1\}&\!\!:\!\!&\left(\begin{array}{c}
I(X^{I}_1:Y^{I}|Q^I)+I(X^{II}_1:Y^{II}|Q^{II})\\
I(X^{I}_2:Y^{I}|X^{I}_1,Q^I)+I(X^{II}_2:Y^{II}|X^{II}_1,Q^{II})
\end{array}\right).
\end{eqnarray}
Using the chain rule,
we obtain that for a given ordered choice $\pi$, rates achieved in vertex $v(\pi)$ are:
\begin{eqnarray}
R_{\pi_i}&=&I(X^I_{\pi_i}:Y^I|X^I_{\pi_{i+1}},\ldots,X^I_{\pi_n},Q^I)+\\
&&I(X^{II}_{\pi_i}:Y^{II}|X^{II}_{\pi_{i+1}},\ldots,X^{II}_{\pi_n},Q^{II})\nonumber.
\end{eqnarray}
and  can be viewed as a sum of the vectors of rates $u(\pi)$ and $w(\pi)$ with entries:
\begin{eqnarray}
R^I_{\pi_i}&=&I(X^I_{\pi_i}:Y^I|X^I_{\pi_{i+1}},\ldots,X^I_{\pi_n},Q^I)\\
R^{II}_{\pi_i}&=&I(X^{II}_{\pi_i}:Y^{II}|X^{II}_{\pi_{i+1}},\ldots,X^{II}_{\pi_n},Q^{II})\nonumber,
\end{eqnarray}
which in an obvious way belong to fixed probability capacity regions $\tilde{\mathcal{R}}_{I}$ and $\tilde{\mathcal{R}}_{II}$ respectively. This completes the proof.


The proof for $n$ MACs can be obtained through induction of the above
proof. Indeed it suffices to divide the set of $n$ MACs into two MACs
-- one composite MAC consisting of $n-1$ MACs and a second channel
consisting of the remaining MAC, and apply the above solution to them.

\begin{acknowledgments}
The work was supported by the EU Commission through the QESSENCE project and by
the Polish Ministry of Science and Higher Education through Grant No. NN202231937. Part of the work was done in Quantum Information Centre of Gdansk.
J.K. was also supported by EU project NAMEQUAM.
\end{acknowledgments}


\begin{thebibliography}{0}
\bibitem{same_input_number_channel_extension}
Note that the assumption that each channel has the same number
of inputs does not in any way influence the generality of the theorem,
since  $m$ can be chosen to be the maximal
number of senders allowed among all considered channels and the
definition of the other channels can be extended simply in the following manner
$p(y|x_A,x_B,\ldots,x_m)=p(y|x_A,x_B,\ldots,x_w)$ if $w<m$.
\bibitem{qeffexp}
Note that we are working in the bright states regime. We do not need single photon resolution. In those conditions queantum efficiency on the level $\eta=99\%$ is avaible e.g. PIN photodiodes.
\end{thebibliography}
\end{document}